\documentclass[12pt,preprint]{aastex}
\usepackage{amssymb}
\usepackage{epsfig}
\usepackage{natbib}
\usepackage{graphicx}
\usepackage{color}
\usepackage{tabularx}
\usepackage{epsfig}
\usepackage{rotating}
\usepackage{multirow}
\usepackage{subfigure}
\usepackage{ulem}
\usepackage{longtable}

\newcommand       \Ks           {{K_{\rm s}}}

\shorttitle{WISE Catalog of Periodic Variable Stars}
\shortauthors{X. Chen et al.}

\begin{document}

\title{Wide-field Infrared Survey Explorer (WISE) Catalog of Periodic Variable Stars}

\author{Xiaodian Chen\altaffilmark{1}, 
Shu Wang\altaffilmark{2}, 
Licai Deng\altaffilmark{1}, 
Richard de Grijs\altaffilmark{3,4},
and Ming Yang\altaffilmark{5}
}
\altaffiltext{1}{Key Laboratory for Optical Astronomy, National
  Astronomical Observatories, Chinese Academy of Sciences, 20A Datun
  Road, Chaoyang District, Beijing 100012, China;
  chenxiaodian@nao.cas.cn}
\altaffiltext{2}{Kavli Institute for Astronomy \& Astrophysics, Peking
  University, Yi He Yuan Lu 5, Hai Dian District, Beijing 100871,
  China; shuwang@pku.edu.cn}  
\altaffiltext{3}{Department of Physics and Astronomy, Macquarie
  University, Balaclava Road, Sydney, NSW 2109, Australia}
\altaffiltext{4}{International Space Science Institute--Beijing, 1
  Nanertiao, Zhongguancun, Hai Dian District, Beijing 100190, China}
\altaffiltext{5}{IAASARS, National Observatory of Athens, Vas. Pavlou
  \& I. Metaxa, Penteli 15236, Greece}

\begin{abstract}
We have compiled the first all-sky mid-infrared variable-star catalog
based on Wide-field Infrared Survey Explorer (WISE) five-year survey
data. Requiring more than 100 detections for a given object, 50,282
 carefully and robustly selected periodic variables are
discovered, of which 34,769 (69\%) are new. Most are located in the
Galactic plane and near the equatorial poles. A method to classify
variables based on their mid-infrared light curves is established
using known variable types in the General Catalog of Variable
Stars. Careful classification of the new variables results in a tally
of 21,427 new EW-type eclipsing binaries, 5654 EA-type eclipsing
binaries, 1312 Cepheids, and 1231 RR Lyraes. By comparison with known
variables available in the literature, we estimate that the
misclassification rate is 5\% and 10\% for short- and long-period
variables, respectively. A detailed comparison of the types, periods,
and amplitudes with variables in the Catalina catalog shows that the
independently obtained classifications parameters are in excellent
agreement. This enlarged sample of variable stars will not only be
helpful to study Galactic structure and extinction properties, they
can also be used to constrain stellar evolution theory and as
potential candidates for the {\sl James Webb Space Telescope}.
\end{abstract}
\keywords{catalogs --- stars: variables: general --- infrared: stars
  --- distance scale --- Galaxy: structure --- Galaxy: disk}

\section{Introduction}

Periodic variable stars exhibit regular or semi-regular luminosity
variations. They are usually discovered through time-domain
surveys. The period is usually related to the object's physical or
geometric properties. The masses, luminosities, radii, and ages of
variable stars can be partially evaluated on the basis of their
periods. Therefore, variables are important objects not only to study
Galactic structure, but also for constraining stellar evolution. Among
the wide diversity of variable stars, classical Cepheids are the most
important distance tracers used to establish the cosmological distance
scale.
 
Before the 1990s, many thousands of periodic variable stars were found
and studied, often in individual papers. After the 1990s, large-scale
surveys became available and the number of periodic variable stars
increased rapidly. The Massive Compact Halo Object survey
\citep[MACHO;][]{Alcock93} found some 20,000 variable stars in the
Large Magellanic Cloud (LMC). Over the course of more than 20 years,
the Optical Gravitational Lensing Experiment (OGLE), releases I--IV
\citep{Udalski94, Udalski15}, detected more than 400,000 variable
stars in the Magellanic Clouds and the Galactic bulge. A Fourier
method was developed to classify different types of variables
\citep{Soszynski08a, Soszynski08b, Soszynski09}. This survey directly
led to a better understanding of the distance to and structure of the
LMC \citep{Pietrzynski13, Inno16}.

The All Sky Automated Survey \citep[ASAS;][]{Pojmanski97} was the
first survey to cover almost the entire sky, probing objects down to
$V \sim 14$ mag. The resulting catalog contained of order 10,000
eclipsing binaries and 8000 periodic pulsating stars
\citep{Pojmanski05}. Similar surveys in the $R$ band were the Robotic
Optical Transient Search Experiment \citep[ROTSE;][]{Akerlof00} and,
subsequently, the Northern Sky Variability Survey
\citep[NSVS;][]{Wozniak04}; \citet{Hoffman09} identified 4659 periodic
variable objects in the NSVS. Recently, near-Earth object (NEO)
surveys have provided new opportunities to find additional periodic
variables. \citet{Sesar11} and \citet{Palaversa13} found and
classified 7000 variables in the course of the Lincoln Near-Earth
Asteroid Research (LINEAR). Significant progress was made in the
context of the Catalina Sky Survey's Northern \citep{Drake14} and
Southern \citep{Drake17} catalogs, which identified some 110,000
periodic variable stars with magnitudes down to $V \sim 20$ mag. This
massively increased the sample of eclipsing binaries and RR Lyrae,
which has resulted in a better understanding of both the solar
neighborhood and the structure of the Galactic halo \citep{Drake13,
  Chen18a}. Nevertheless, all of these surveys were conducted at
optical wavelengths, which prevented detailed studies of the heavily
reddened Galactic disk. The VISTA Variables in V\'{i}a L\'{a}ctea
(VVV) team conducted a near-infrared (NIR) $\Ks$-band multi-epoch
survey of the Galactic bulge and southern disk. They found 404 RR
Lyrae in the southern Galactic plane \citep{Minniti17}.

The study of variables in the infrared has great potential given the
current availability of the {\sl Spitzer Space Telescope} and the
Wide-field Infrared Survey Explorer (WISE), and also in view of the
soon to be launched {\sl James Webb Space Telescope} (JWST). Studies
of variables in the infrared view can benefit our understanding of
infrared extinction, since the universality (or otherwise) of the
infrared extinction law is still being debated
\citep[e.g.][]{Matsunaga18}. Studies using the red clump as a
diagnostic tool imply a relatively universal extinction law in the NIR
but a variable mid-infrared (MIR) extinction law \citep{Wang14,
  Zasowski09}. Compared to using the red clump, samples of periodic
variables can be isolated and identified much better. Cepheids and RR
Lyrae, well-understood secondary distance tracers, can also be used to
study the structure of and distances to the Galactic arms, bulge, bar,
and center. \citet{Matsunaga11} first detected three classical
Cepheids in the Galactic Center's nuclear stellar disk based on NIR
photometry. \citet{Feast14} traced the flaring of the outer disk based
on five classical Cepheids using both optical and NIR photometry. The
spiral arm are expected to be delineated by these variables. In
addition, the number of stars accompanied by dust shells or disks will
increase prominently if selections are based on infrared
observations. This provides additional opportunities to better
understand the physical properties of Miras, semi-regular variables,
Be stars, and many other types of variables.

In this paper, we collect the five-year WISE data to detect periodic
variables across the entire sky. More than 50,000 high-confidence
variables are found, filling the gaps in the Galactic plane. An
all-sky variable census is achieved down to a magnitude $W1\sim14$ mag
($G\sim16$ mag). Light-curve and photometric analyses are used to
classify the variables. More than 34,000 new variables are found. The
data are described in Section 2. The methods adopted to identify and
classify variables are discussed in Sections 3 and 4,
respectively. Our new variable catalog and a comparison of its
properties with previously published parameters are included in
Section 5. Section 6 presents the light curves of different variables
and Section 7 concludes the paper.

\section{WISE multi-epoch data}

WISE is a 40 cm (diameter) space telescope with a $47' \times 47'$
field of view. It was designed to conduct an all-sky survey in four
MIR bands, $W1$ (3.35 $\mu$m), $W2$ (4.60 $\mu$m), $W3$ (11.56
$\mu$m), and $W4$ (22.09 $\mu$m). WISE began to operate in survey mode
in 2010 January; it covers the full sky once every half a year
\citep{Wright10}. The solid hydrogen cryostat was depleted in 2010
September, upon which a four-month NEOWISE Post-Cryogenic Mission
\citep{Mainzer11} continued to accomplish the second all-sky
coverage. The data resulting from these two epochs were packaged as
the ALLWISE Data Release, characterized by an enhanced accuracy and
sensitivity compared to the WISE All-Sky Data Release. Next, WISE
entered into a hibernating state for two years and was reactivated in
2013 October. WISE operations are currently continuing with a mission
called the `near-Earth object WISE reactivation (NEOWISE-R) mission'.
At present, four years of NEOWISE-R data have been released.

Without cryogen, NEOWISE-R only performs photometry in two bands, $W1$
and $W2$. Therefore, by combining ALLWISE and NEOWISE-R, five whole
years and photometric data covering at least 10 epochs are available
for each object. During each epoch, some 10--20 images were taken of
any one field with intervals of 0.066 or 0.132 days; the total
exposure time for a given epoch ranged between 1.1 and 1.4 days. The
majority of stars have more than 100 detections and variables with
periods in the range of 0.14--10 days can be identified
adequately. The zeropoint differences between ALLWISE and NEOWISE-R
were discussed in detail by \citet{Mainzer11}; a systematic difference
of $\sim$0.01 mag and a statistical scatter of 0.03--0.04 mag for
$8<W1<14$ mag and $7<W2<13$ mag were found. The angular resolution of
WISE, as a function of passband at increasing wavelengths is $6.1'',
6.4'', 6.5''$, and $12.0''$, which is much lower than that of the Two
Micron All Sky Survey \citep[2MASS;][]{Cutri03}. To reduce blending,
2MASS photometry was used to assist. The combination of NIR and MIR
photometry was also used to select variable stars exhibiting infrared
excesses.

The ALLWISE and NEOWISE-R catalogs contain about 100 billion rows of
data for 0.7 billion individual objects. This is simply too large a
number to access the entire database and perform the
analysis. Therefore, we only focused on objects that are
high-probability variables, i.e., with AllWISE Source Catalog keyword
`var\_flg' = 6, 7, 8, or 9. Although this selection could potentially
omit a small number of variables, it is indeed helpful to avoid the
detection of pseudo-light variations. The total number of candidates
is 2.7 million. Data pertaining to multi-epoch photometry was
collected by selecting angular distances of less than $1''$ for any
given object. Detections with `null' magnitude uncertainties were
excluded, since their magnitudes were either not measurable or 95\%
confidence-interval upper limits. This selection criterion also
excludes photometry with poor signal-to-noise ratios, $snr=$ `null,'
and poor reduced $\chi^2$ resulting from the profile fits, $rchi2=$
`null.'  $qi\_fact>0, saa\_sep>0, moon\_masked=0$, and $qual\_frame>0$
were adopted to exclude other types of contamination or biases:
$qual\_fact$ is the frame quality score, which ranges from 0 to 10 (0
represents the lowest quality, i.e., sources that are spurious
detections due to noise noise, or objects affected by transient events
or scattered light), $qi\_fact$ is the image quality score (where
$qi\_fact=0$ represents the lowest quality), and $saa\_sep$ is the
distance to the boundary of the South Atlantic Anomaly, in degrees
($saa\_sep>0$ means that WISE is on the outside). Finally,
$moon\_masked=0$ denotes that the frame is not located in the
moon-masked area.

\section{Identification of periodic variables}

To identify periodic variables, the Lomb--Scargle periodogram
\citep{Lomb76, Scargle82} method was adopted. The input frequency
ranged from 0.01 to 7 day$^{-1}$ with a step length of 0.0001
day$^{-1}$, so it covered periods between 0.143 and 10 days, with an
ideal accuracy of better than 0.01\%. Periods were derived from the
frequency associated with the maximum power spectral density
(PSD). The standard deviation of the frequencies, $\sigma_{50\%}$, and
$\sigma_{95\%}$ were estimated based on PSDs greater than 50\% and
95\% of the maximum PSD, respectively. The preliminary period
uncertainty, $\sigma_1(P)$, was estimated based on the
latter. $\sigma_{50\%}<0.21$ and $\sigma_1(P)/P<0.20$ were used to
select variable candidates; $\sigma_{50\%}=0.21$ is the boundary where
the peak becomes flat. This selection resulted in a rate of false
noise of less than 10\%. Then, we adopted a fourth-order Fourier
analysis $\displaystyle{f =a_0 + \sum_{i=1}^4a_i\cos(2\pi
  it/P+\phi_i)}$ to fit each light curve, and adjusted $R$-square
($R^2>0.7$) and root-mean-squared error (RMSE $<$ 0.05) were combined
to select candidates. The phase idleness rate was estimated using a
bin size of 0.01, and objects with idleness rates greater than 50\%
were excluded. In addition, objects with poorly fitting light curves
were excluded through higher-order amplitude selection, $a_3>0.08$ mag
or $a_4>0.08$ mag. After application of these selection criteria,
68,034 candidates remained.

Based on a visual check of the light curves of some candidate
variables, it became clear that major problems were caused by the
imhomogeneous distribution of the data points. This would lead to the
derivation of incorrect periods. The data were divided into five bins,
by phase, without enforcing a fixed starting point, and the number
probability, mean deviation, and scatter around the best-fitting
Fourier line in each bin were estimated.  Note that the mean deviation
and scatter were given in units of total amplitude. These three
parameters were used to exclude objects located outside the $3\sigma$
boundaries for any given selection choice and outside $1\sigma$ for
all three selection criteria. The objects thus excluded were flagged
as suspected variables, since their period could likely be improved
with larger numbers of detections.

\begin{figure}[h!]
\centering
\vspace{-0.0in}
\includegraphics[angle=0,width=180mm]{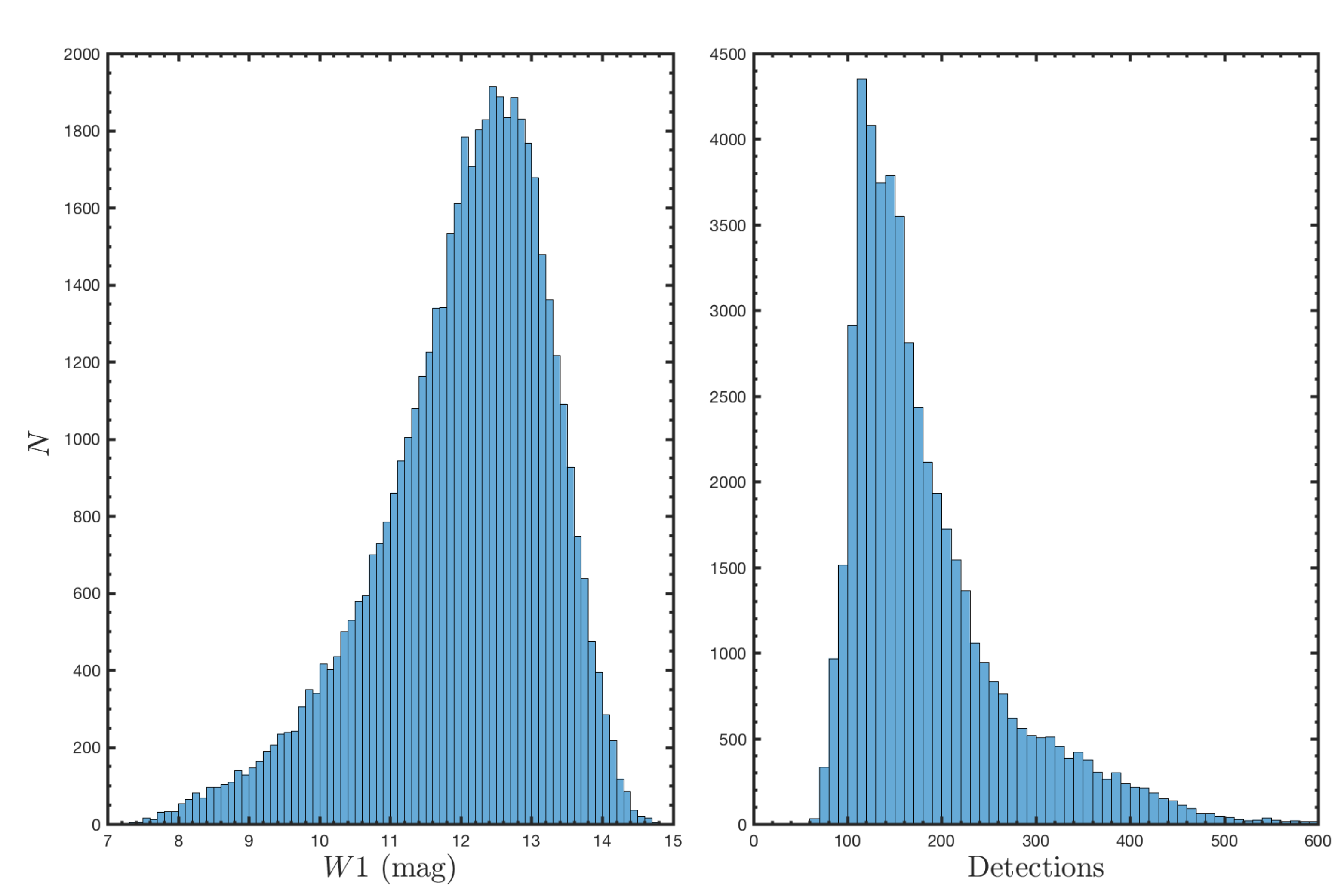}
\vspace{-0.0in}
\caption{\label{mag} Number distribution of (left) $W1$ magnitudes and
  (right) $W1$ detections for our 50,296 variables.}
\end{figure}

\begin{figure}[h!]
\centering
\vspace{-0.0in}
\includegraphics[angle=0,width=160mm]{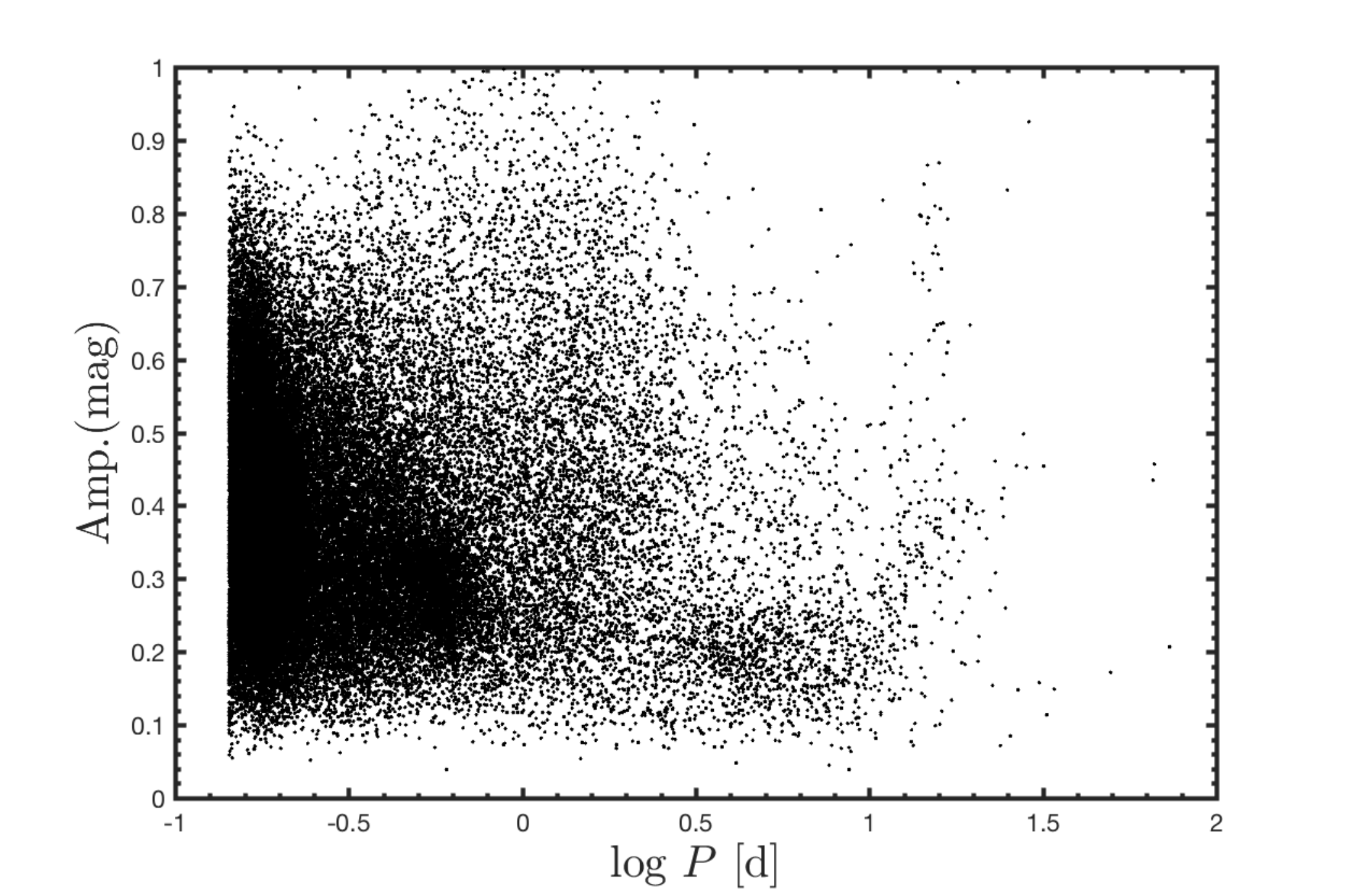}
\vspace{-0.0in}
\caption{\label{ampper} Amplitude versus period diagram for the 50,296
  variables.}
\end{figure}

The final selection step aimed at excluding objects with unstable
periods, especially among long-period objects. The best periods were
determined using the four- and five-year data separately. Objects with
period differences in excess of 10\% were also excluded as suspected
variables. The remaining objects comprised our sample of
high-confidence variables. The period difference was treated as a
second period uncertainty, $\sigma_2(P)$, and the final period
uncertainty was the larger of $\sigma_1(P)$ and $\sigma_2(P)$. The
total number of variables thus obtained was 50,296. The $W1$ magnitude
distribution of these variables is shown in the left-hand panel of
Figure \ref{mag}. Most are in the magnitude range $8<W1<14$ mag, where
the zeropoint differences between ALLWISE and NEOWISE-R are
negligible. The right-hand panel of Figure \ref{mag} shows the
$W1$-band detection numbers, with a upper limit of 600 (489 variables
have more than 600 detections). We can infer that only 2853 variables
have fewer than 100 detections, which means that our final sample of
50,296 variables is well-covered in the $W1$ band. Figure \ref{ampper}
is the amplitude versus period diagram, which is used to detect false
positives. False signals due to sampling patterns would produce a
random distribution in amplitude for a given period
\citep{Drake14}. Based on Figure \ref{ampper}, we assert that no such
false signals are detected among our 50,296 variables.

\section{Classification of periodic variables}

\begin{figure}[h!]
\centering
\vspace{-0.0in}
\includegraphics[angle=0,width=180mm]{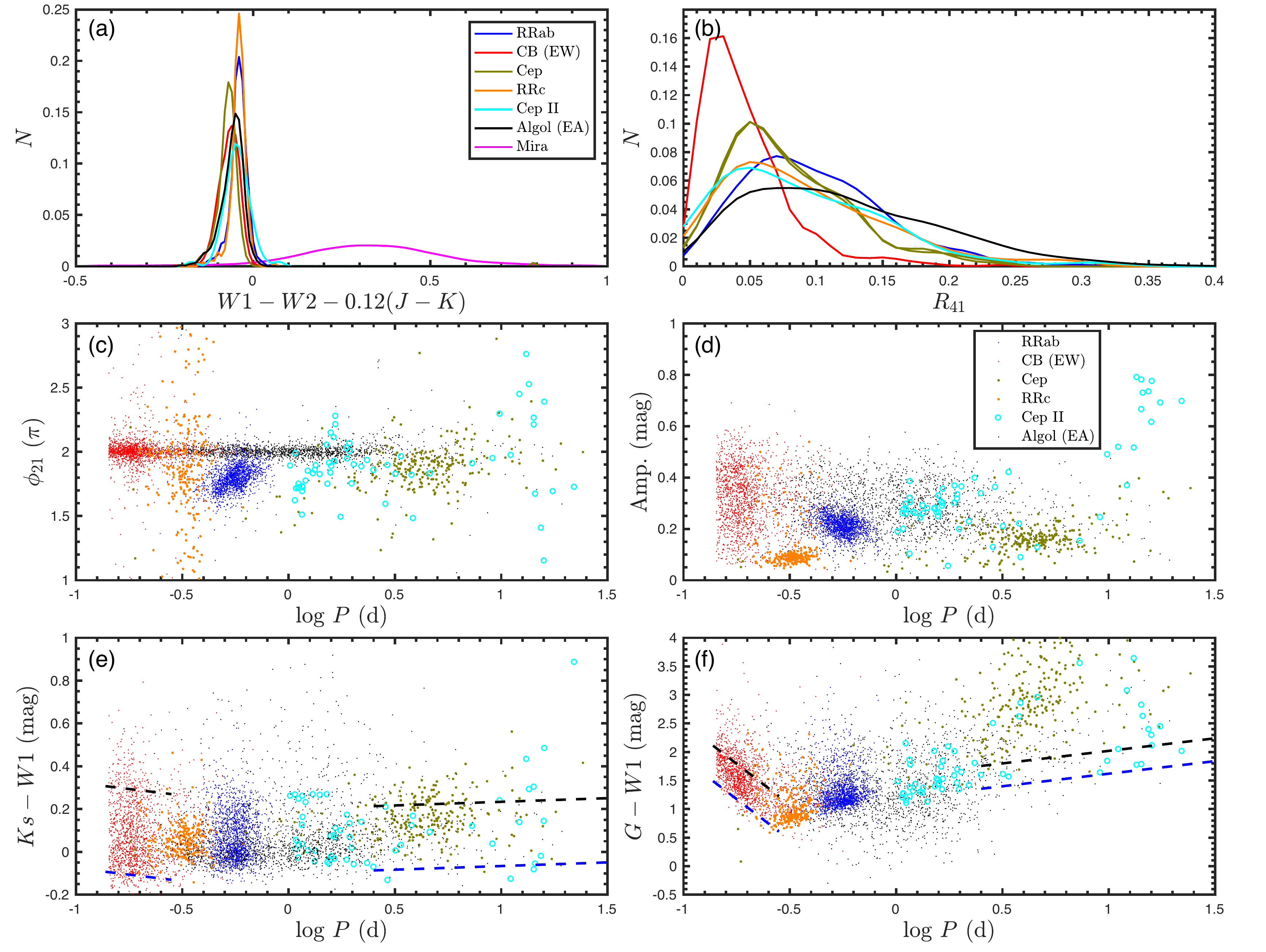}
\vspace{-0.0in}
\caption{\label{cl} The different colors and symbols denote different
  types of variables. Blue: RRab; red: EWs; light green: Cep-Is;
  orange: RRc; cyan: Cep-IIs; black: EAs; magenta: Miras. The six
  panels show the parameters used to classify our variables: infrared
  excess index (a), period (b, d, e, f), phase difference (b, c),
  amplitude ratio (c), full amplitude (d), $\Ks-W1$ color (e), and
  $G-W1$ color (f). The black and blue dashed lines in panels (e) and
  (f) are the intrinsic colors of contact binaries and classical
  Cepheids, respectively \citep{Gaia17, Chen18a, Wang18}. }
\end{figure}

To classify the types of periodic variables, their colors, periods,
and shapes of the light curves were used. We tested our
classifications using variables with known classifications from the
General Catalog of Variable Stars (GCVS). In the magnitude detection
range of WISE, 80\%, 60\%, 49\%, 46\%, 43\%, and 42\% of EW-type
eclipsing binaries (EWs), classical Cepheids (Cep-Is), Type-ab RR
Lyrae (RRab), EA-type eclipsing binaries (EAs), Type-c RR Lyrae (RRc),
and Type II Cepheids (Cep-IIs) were confirmed. We did not detect Mira
variables because of their long periods. Therefore, we used GCVS Miras
to analyze the infrared excesses of our sample Miras. To guide our
classification, Figure \ref{cl} shows the distributions of variables
using different parameters.

{\sl Infrared excess}. Since the variable stars were selected in the
infrared, a number of objects are expected to have intrinsic infrared
excesses owing to the presence of circumstellar dust disks or
shells. To exclude these objects from normal stars, \citet{Flaherty07}
adopted selection criteria $[3.6]-[4.5]<0.6$ mag and $[5.8]-[8.0]<0.2$
mag based on {\sl Spitzer} photometry. The effective wavelengths of
the WISE $W1$ and $W2$ bands are close to those of the {\sl Spitzer}
$[3.6]$ and $[4.5]$ filters. Because of the low accuracy of our $W3$
and $W4$ photometry, we adopted as selection criterion
$W1-W2<0.12\times(J-\Ks)$. This is somewhat stricter than
\citet{Flaherty07}'s $[3.6]-[4.5]<0.6$ mag, since most of the
variables are characterized by $J-\Ks<5.0$ mag. Figure \ref{cl}a shows
the Gaussian kernel smooth density distributions of several types of
variables. EAs, EWs, RRab, RRc, and Cep-Is all satisfy the criterion,
while the evolved Miras entirely exceed it. Variables with infrared
excess include not only Miras, but also Be stars, young stellar
objects, and a small fraction of Cep-II.

{\sl Periods}. Periods of variables are in part related to their
luminosities and masses, so different types of variables have
characteristic periods. In Figure \ref{cl}d, e, and f, different
colors and symbols denote the period distribution of these variables
(for eclipsing binaries we provide half periods). From short to long
periods, we encounter EWs, RRc, RRab, EAs, Cep-Is and Cep-IIs.
Although EW-type eclipsing binaries are found with periods ranging
from 0.19 days to a few dozen days, the vast majority have periods of
less than 1 day. EA-type eclipsing binaries also span a wide period
range. However, the majority have periods longer than 1 day. The
period ranges of RRc and RRab are relatively concentrated:
$0.16<P<0.50$ days and $0.32<P<1.00$ days, respectively
\citep{Clementini18}. Cep-IIs were divided into three subtypes: BL
Herculis (BL Her), W Virginis (W Vir), and RV Tauri (RV Tau), with the
corresponding period ranges $1.0\leq P<4.0$ days, $4.0\leq P<20.0$
days, and $P\geq 20.0$ days \citep{Soszynski08b}. In this paper, we
could only detect BL Her and W Vir variables. The periods of Cep-Is
are usually longer than 2.24 days \citep{Gaia17}.

{\sl Colors}. Since the distances to our variables are usually
unknown, we adopted their intrinsic colors instead of their absolute
magnitudes. EW-type eclipsing binaries are redder than main-sequence
stars, and they become bluer for increasing orbital periods. RRab and
RRc are bluer still and concentrated in a narrow color range
\citep[see also][]{Drake14}. They become slightly redder for
increasing periods, which is caused by changes in the bolometric
correction. Cep-Is also become redder with increasing of period. Since
Galactic Cep-Is are usually located in the heavily reddened mid-plane,
they are often associated with significant extinction and
reddening. Figure \ref{cl}e and f shows the distributions of our
variables in $\Ks-W1$ and $G-W1$ color vs. $\log P$ space. The
intrinsic colors of contact binaries and Cepheids \citep{Gaia17,
  Chen18a, Wang18} are shown as the black and blue dashed lines,
respectively, for reference. Note that Cep-IIs exhibit a wide spread
in $\Ks-W1$ color, which is the result of their intrinsic infrared
excesses rather than foreground reddening. For variables affected by
low reddening, EWs are redder than $G-W1=1.0$ mag, RRcs are bluer than
$G-W1=1.0$ mag, while RRab are found in the range of $1.0<G-W1<1.5$
mag. However, for variables affected by significant reddening, no
color selection criteria are available, since $\Ks-W1$ is both
reddening and temperature insensitive.

{\sl Light curves}. The shape of the light curve is of key importance
to distinguish among variables. In Figure \ref{cl}c, eclipsing
binaries have more symmetric light curves, which is reflected in the
phase difference $\phi_{21}=\phi_{2}-2\phi_{1}$. This may deviate from
$0$ or $2\pi$ (we adopt $2\pi$ uniformly) because of low photometric
accuracy, low amplitude, or complexity in the light curve. In the WISE
$W1$ band, 85\% of EWs and 93\% of EAs in the GCVS have phase
differences in the range $|\phi_{21}-2\pi|<0.1\pi$. On the other hand,
pulsating stars exhibit asymmetric light curves, which become more
symmetric from optical to infrared wavelengths. Nevertheless, in the
$W1$ band the asymmetry is still recognizable. The phase differences
of RRab increase with period in the range of $[1.5-2.0]\pi$. RRc have
a random distribution of $\phi_{21}$ because of their low
amplitudes. Cep-Is have a phase difference in the range
$\phi_{21}=[1.5-2.1]\pi$ for $2<P<10$ days and a random distribution
around $P=10$ days. The global trend is similar to that in the
$I$-band Fourier analysis based on the OGLE survey
\citep{Soszynski08a}. Our limited sample of Cep-IIs are more scattered
than the classical Cepheids. The amplitude ratio $R_{41}=a_4/a_1$
could be used to distinguish EAs from other types of variables, since
Cep-Is and EWs exhibit a cut-off at $R_{41}=0.25$ and $0.20$,
respectively (see Figure \ref{cl}b). The amplitude is the other
characteristic parameter. Figure \ref{cl}d shows the distribution of
the best-fitting amplitudes. Note that the amplitudes were only used
to classify the variables; the optimal amplitudes were redetermined
after the classification step (see Section \ref{subamp}). Eclipsing
binaries show a random distribution between 0.05 mag and 0.75 mag
while pulsating stars are characterized by a more concentrated
distribution. RRc have the lowest amplitudes (mostly less than 0.15
mag); the amplitude distribution of RRab and RRc is comparable to that
of \citet{Gavrilchenko14}. The amplitude of Cep-Is is usually smaller
than 0.3 mag, while Cep-IIs show a wide distribution.

\subsection{Variable types}

\begin{table}[h!]
\vspace{-0.0in}
\begin{center}
\caption{\label{t1}Criteria adopted to classify different variables}
\vspace{0.15in}
\begin{tabular}{ll}
\hline
\hline
                             Type& Selection criteria \\   
\hline                             
infrared excess variable              & $W12JK>0$, $P>2.24$ d                                                     \\     
RR             & $0.316<P<1.0$ d, $\phi_{21}<1.9\pi$                                         \\   
Cep-I             & $W12JK<0$, $P>2.24$ d, $\rm{Amp.}<0.5$ mag, $1.6\pi<\phi_{21}<2.1\pi$             \\  
Cep-I/Cep-II            & $W12JK<0$, $P>2.24$ d, $\rm{Amp.}>0.5$ mag                                      \\     
                  & $W12JK<0$, $P>2.24$ d, $\rm{Amp.}<0.5$ mag, $|\phi_{21}-1.85\pi|>0.25\pi$          \\ 
Cep-II/ACep/Cep-I    & $1.0<P<2.24$ d, $\phi_{21}<1.9\pi$                                  \\  
Cep-I/EA   &   $P>2.24$ d, $W12JK<0$,  $|\phi_{21}-2.0\pi|<0.12\pi$,  $0.15<R_{31}<0.22$ \\                   
    Eclipsing binary & $0.16<P<0.316$ d, $\rm{Amp.}\geq0.15$ mag                                       \\     
                 & $P<0.16$ d                                                                \\     
                  & $0.316<P<1.0$ d, $\phi_{21}>2.1\pi$               \\
                 & $P<2.24$ d, $|\phi_{21}-2.0\pi|<0.1\pi$                                      \\   
                 & $P>2.24$ d, $W12JK<0$,  $|\phi_{21}-2.0\pi|<0.12\pi$,  ($R_{21}>0.4$  or $R_{41}>0.15$)   \\   
Misc           & $0.16<P<0.316$ d, $\rm{Amp.}<0.15$ mag                                          \\ 
                   & $1.0<P<2.24$ d, $\phi_{21}>2.1\pi$                                  \\   
\hline                                                                                     
EW               & $a_4<a_2(a_2+0.125)$                                                      \\
EA               & $a_4>a_2(a_2+0.375)$                                                      \\     
                 & $a_2(a_2+0.125)<a_4<a_2(a_2+0.375)$,  $P>1.0$ d                           \\     
EW/EA            & $a_2(a_2+0.125)<a_4<a_2(a_2+0.375)$,  $P<0.5$ d                           \\     
EA/RR/EW            & $a_2(a_2+0.125)<a_4<a_2(a_2+0.375)$,  $0.5<P<1.0$ d                           \\  

\hline
\end{tabular}
\end{center}
\end{table}

\begin{figure}[h!]
\centering
\vspace{-0.0in}
\includegraphics[angle=0,width=120mm]{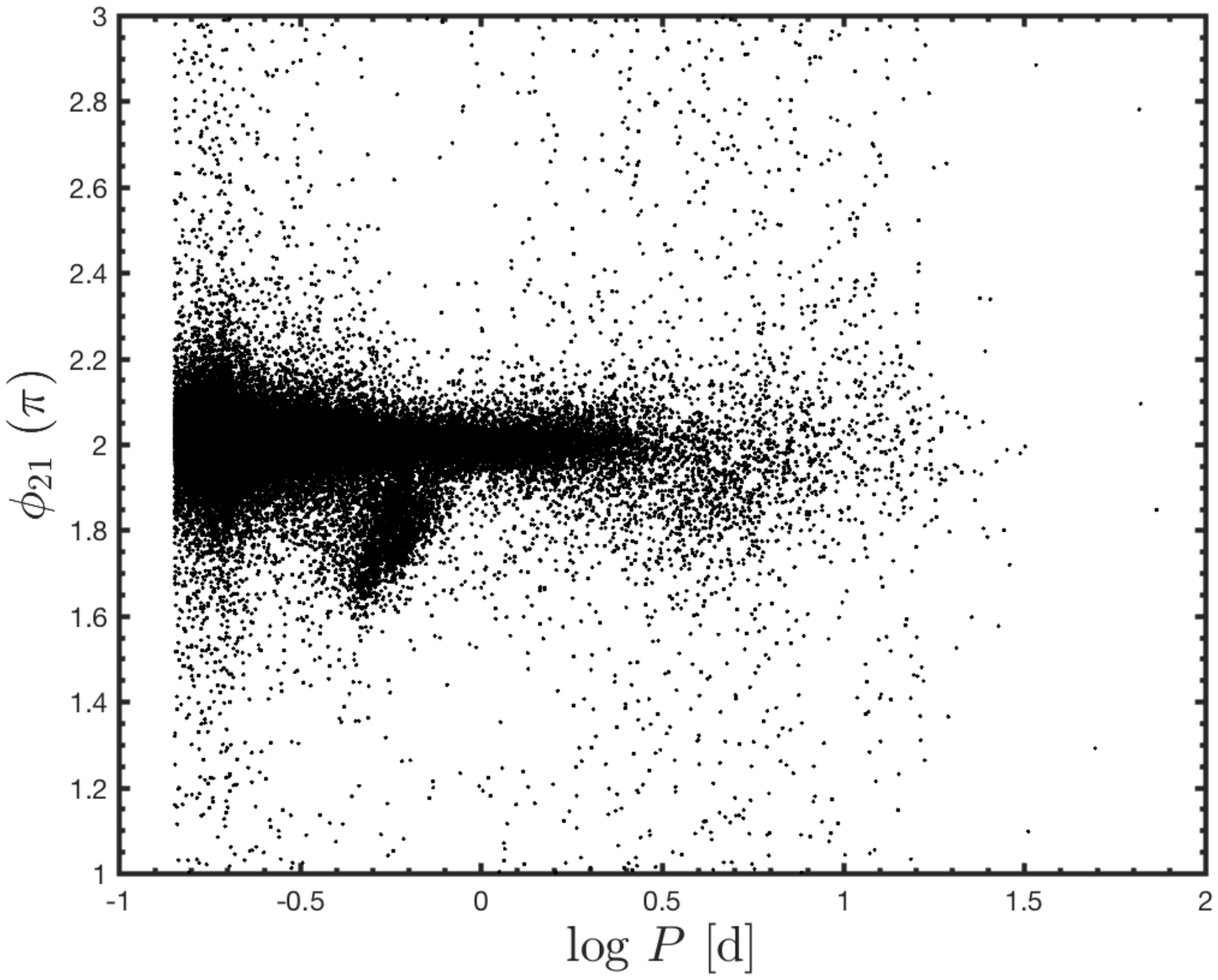}
\vspace{-0.0in}
\caption{\label{allf} $\phi_{21}-\log P$ diagram defined by our 50,296
  variables. The distribution is similar to Figure
  \ref{cl}c. Eclipsing binaries are most obvious around
  $\phi_{21}=2\pi$. RRab and Cep-Is are found in two overdense
  regions, $-0.5<\log P<0.0$ and $\log P>0.5$, respectively (see also
  Figure \ref{cl}c.}
\end{figure}

Employing the combination of these properties, EW- and EA-type
eclipsing binaries, RR Lyrae, Cep-Is, Cep-IIs, and variables
exhibiting an infrared excess could be classified (see Figure
\ref{allf}). Note that we did not use any color information, since
many of our sample objects were reddened, so that intrinsic colors
could not be determined. Table \ref{t1} lists the criteria adopted to
classify the different types of variables. Infrared-excess variables
were selected only for periods longer than 2.24 days, using
$W12JK=W1-W2-0.12(J-\Ks)>0$. Most were Miras and semi-regular
variables (SRs), although small fractions of Cep-IIs, Be stars, and
young stellar objects contaminated the selection. We were unable to
detect any Miras or SRs with accurate periods, since the periods of
these types of objects far exceed 10 days. Therefore, all such
candidates were added to the suspected variables catalog.

Given the more symmetric MIR light curves, the boundary between RR
Lyrae and eclipsing binaries was less obvious. Therefore, we only
selected RR Lyrae exhibiting obvious asymmetries, $\phi_{21}<1.9\pi$,
in the period range attributed to RR Lyrae; the remaining objects were
classified as `EA/RR/EW.' Compared with RRab, RRc have shorter periods
and smaller amplitudes. They could not be easily classified. These
latter objects are found mixed in with the EW, RR, and miscellaneous
variables (`Misc'). Cep-Is and Cep-IIs could be distinguished
approximately by their amplitude and phase differences. EA-type
objects were excluded from the Cep-I/II bin based on their larger
amplitude ratio. The light curves of Cepheids were checked by eye to
exclude long-period eclipsing binaries, because eclipsing binaries
spend less time in the actual eclipse. Some 5--10\% of the Cepheid
subsample are likely eclipsing binaries. Anomalous Cepheids (ACeps)
are mixed with Cep-Is and Cep-IIs in the period range $1.0<P<2.24$
days. The remaining variables were eclipsing binary candidates. Note
that the parameter $P$ used in this case is the half-period of the
eclipsing binaries. To distinguish between EW- and EA-type eclipsing
binaries, we reran our Fourier analysis using the full periods. The
curves $a_4=a_2(a_2+0.125)$ and $a_4=(a_2+0.375)$ were adopted to
refine the EA and EW selection \cite[see also][]{Chen18a}; $a_2$ and
$a_4$ are the amplitudes of the second- and fourth-order Fourier
series.  Objects located near the boundary were classified as `EW/EA.'
A few hundred candidates showing asymmetric light curves across their
full periods were reclassified as RR Lyrae. EB-type ($\beta$ Lyrae)
eclipsing binaries could not be classified and are mixed with the EA
and EW types \citep[see also][]{Drake17}. Variables such as RS CVn,
$\delta$ Scuti objects, and rotating stars could not be classified
since their MIR light curves did not provide sufficient characteristic
features. These latter variables are either mixed with the other types
or classified as Misc.

\begin{table}[h!]
\tiny
\vspace{-0.0in}
\begin{center}
\caption{\label{t4}WISE variables catalog \tablenotemark{a}.}
\vspace{0.15in}
\begin{tabular}{lccccccccccccccc}
\hline
\hline
ID  & R.A.     & Decl.  & $<W1>$ & Period & err$_P/P$ & Num & Amp.$_{\rm{F10}}$ & Amp.$_{10}$ & Type & $R_{21}$ & $\phi_{21}$ &...\\                           
           & $^\circ$ & $^\circ$ & mag    & days    &           &     &  mag              & mag         &          & &$\pi$            &... \\ 
\hline                         
WISEJ094812.4$+$093448     &  147.05182  &    9.58021 &  11.667  &  0.6664177  & 0.00001    &  115  &  0.278  &  0.257 &           EA &  0.504  &  2.041  &  ...\\
WISEJ130034.1$-$211735     &  195.14238  &  $-$21.29326 &  11.306  &  0.6489507  & 0.00179    &   95  &  0.320  &  0.300 &        EW/EA &  0.319  &  2.035  &  ...\\
WISEJ125656.5$+$752821     &  194.23544  &   75.47264 &   8.298  &  1.6809747  & 0.00464    &  309  &  0.263  &  0.265 &           EW &  0.103  &  2.139  &  ...\\
WISEJ162004.0$+$451300     &  245.01694  &   45.21670 &   8.249  &  0.9836457  & 0.00269    &  269  &  0.163  &  0.169 &           EW &  0.066  &  2.223  &  ...\\
WISEJ164755.1$+$351756     &  251.97993  &   35.29889 &  12.118  &  0.6838987  & 0.00190    &  225  &  0.244  &  0.259 &           EW &  0.186  &  2.076  &  ...\\
WISEJ141332.8$+$031837     &  213.38667  &    3.31038 &   9.253  &  0.9079302  & 0.00002    &  114  &  0.175  &  0.147 &           EW &  0.099  &  2.170  &  ...\\
WISEJ113905.7$+$663017     &  174.77396  &   66.50495 &  11.343  &  1.6719398  & 0.00001    &  275  &  0.121  &  0.746 &           EW &  0.029  &  2.703  &  ...\\
WISEJ095035.2$+$354215     &  147.64679  &   35.70417 &   9.547  &  0.8519414  & 0.00165    &  116  &  0.196  &  0.182 &           EW &  0.056  &  2.457  &  ...\\
WISEJ151428.7$+$142730     &  228.61967  &   14.45854 &  10.535  &  1.0650266  & 0.00293    &  145  &  0.245  &  0.229 &           EW &  0.171  &  2.049  &  ...\\
WISEJ142230.6$+$012758     &  215.62791  &    1.46638 &  10.775  &  0.8127400  & 0.00001    &  129  &  0.204  &  0.210 &           EW &  0.069  &  2.228  &  ...\\
WISEJ145449.0$+$120526     &  223.70424  &   12.09083 &  12.677  &  0.7269519  & 0.00003    &  140  &  0.306  &  0.304 &           EW &  0.230  &  2.103  &  ...\\
WISEJ161622.5$+$243045     &  244.09389  &   24.51251 &  11.799  &  0.7506710  & 0.00001    &  162  &  0.254  &  0.218 &           EA &  0.559  &  2.098  &  ...\\
WISEJ102856.8$-$021733     &  157.23707  &   $-$2.29270 &  11.471  &  1.1148194  & 0.00004    &  133  &  0.296  &  0.268 &     EA/RR/EW &  0.354  &  2.038  &  ...\\
WISEJ102512.0$+$645138     &  156.30015  &   64.86075 &  10.079  &  0.7137239  & 0.00197    &  209  &  0.163  &  0.170 &           EW &  0.086  &  2.288  &  ...\\
WISEJ131817.1$-$123826     &  199.57147  &  $-$12.64065 &  12.707  &  0.8287763  & 0.00230    &  113  &  0.234  &  0.226 &        EW/EA &  0.365  &  2.025  &  ...\\
WISEJ125214.2$+$385630     &  193.05918  &   38.94193 &  10.723  &  0.6424413  & 0.00177    &  147  &  0.289  &  0.268 &           EW &  0.213  &  2.034  &  ...\\
WISEJ120251.4$+$334617     &  180.71442  &   33.77155 &  12.424  &  1.2186550  & 0.00335    &  156  &  0.341  &  0.306 &     EA/RR/EW &  0.391  &  2.072  &  ...\\
WISEJ101258.6$-$125418     &  153.24445  &  $-$12.90523 &  12.288  &  0.8111031  & 0.00157    &  110  &  0.249  &  0.247 &           EW &  0.078  &  2.106  &  ...\\
WISEJ141240.9$-$280107     &  213.17047  &  $-$28.01873 &  12.766  &  0.8297233  & 0.00228    &  126  &  0.237  &  0.251 &           EW &  0.131  &  2.058  &  ...\\
WISEJ095402.2$-$111655     &  148.50936  &  $-$11.28209 &  11.665  &  0.8831864  & 0.00245    &   98  &  0.204  &  0.206 &           EW &  0.049  &  2.019  &  ...\\
WISEJ170410.6$+$274628     &  256.04421  &   27.77450 &  13.833  &  0.7093259  & 0.00195    &  224  &  0.523  &  0.501 &           EW &  0.248  &  2.024  &  ...\\
WISEJ171731.0$+$524901     &  259.37949  &   52.81721 &   8.184  &  0.9313811  & 0.00180    &  418  &  0.210  &  0.223 &           EW &  0.052  &  2.253  &  ...\\
WISEJ122859.8$+$835851     &  187.24920  &   83.98096 &  13.078  &  0.7773740  & 0.00214    &  298  &  0.227  &  0.265 &           EW &  0.118  &  2.048  &  ...\\
WISEJ172508.1$+$371155     &  261.28400  &   37.19883 &  10.688  &  0.4412721  & 0.00242    &  227  &  0.222  &  0.243 &           RR &  0.019  &  1.890  &  ...\\
WISEJ123710.2$-$253930     &  189.29260  &  $-$25.65853 &  13.159  &  0.6405664  & 0.00247    &   99  &  0.319  &  0.325 &           RR &  0.304  &  1.897  &  ...\\
WISEJ092049.3$+$005246     &  140.20578  &    0.87956 &   8.014  &  1.2632250  & 0.00004    &  116  &  0.184  &  0.180 &           EW &  0.051  &  2.016  &  ...\\
WISEJ142750.2$-$203408     &  216.95948  &  $-$20.56915 &  12.344  &  0.7971740  & 0.00439    &  101  &  0.148  &  0.240 &           RR &  0.090  &  1.933  &  ...\\
WISEJ104541.0$-$203634     &  161.42120  &  $-$20.60958 &   8.648  &  0.6751995  & 0.00003    &  102  &  0.170  &  0.168 &           EA &  0.560  &  2.060  &  ...\\
WISEJ154116.9$+$744410     &  235.32058  &   74.73617 &  12.219  &  0.7267838  & 0.00003    &  612  &  0.158  &  0.218 &           EW &  0.079  &  2.080  &  ...\\
WISEJ171456.7$+$585128     &  258.73663  &   58.85789 &  11.724  &  0.8399786  & 0.00231    &  368  &  0.172  &  0.222 &           EW &  0.057  &  2.926  &  ...\\
WISEJ163216.0$+$750624     &  248.06705  &   75.10689 &  11.899  &  0.7972567  & 0.00155    &  335  &  0.141  &  0.160 &           EW &  0.051  &  2.342  &  ...\\
WISEJ151814.3$+$831733     &  229.55985  &   83.29276 &  11.418  &  1.7499784  & 0.00478    &  331  &  0.157  &  0.174 &           EW &  0.156  &  2.167  &  ...\\
WISEJ092206.6$-$103321     &  140.52763  &  $-$10.55590 &  13.094  &  0.6951391  & 0.00001    &  108  &  0.320  &  0.321 &        EW/EA &  0.277  &  2.068  &  ...\\
WISEJ072555.4$+$582631     &  111.48094  &   58.44205 &  12.777  &  0.6412162  & 0.00101    &  146  &  0.290  &  0.286 &           EW &  0.092  &  2.146  &  ...\\
WISEJ180409.2$+$445727     &  271.03869  &   44.95756 &  12.182  &  0.6664398  & 0.00129    &  342  &  0.127  &  0.161 &           EW &  0.091  &  2.235  &  ...\\
WISEJ052330.6$+$875817     &   80.87787  &   87.97140 &  12.968  &  0.8472612  & 0.00002    &  324  &  0.264  &  0.279 &           EW &  0.076  &  2.172  &  ...\\
WISEJ081021.7$+$324011     &  122.59043  &   32.66983 &  12.571  &  0.7071032  & 0.00001    &  106  &  0.349  &  0.336 &        EW/EA &  0.273  &  2.071  &  ...\\
WISEJ170425.6$+$061932     &  256.10676  &    6.32564 &  11.614  &  0.7263457  & 0.00199    &  134  &  0.233  &  0.225 &           EW &  0.077  &  2.026  &  ...\\
WISEJ163508.1$-$003634     &  248.78400  &   $-$0.60956 &  11.627  &  0.6934337  & 0.00134    &  127  &  0.368  &  0.406 &           EW &  0.146  &  2.043  &  ...\\
WISEJ180209.3$+$441950     &  270.53889  &   44.33074 &  12.520  &  1.4118240  & 0.00273    &  359  &  0.222  &  0.256 &           EW &  0.062  &  2.122  &  ...\\
WISEJ083550.2$+$071655     &  128.95934  &    7.28196 &  10.887  &  0.9797850  & 0.00002    &  110  &  0.271  &  0.273 &           EW &  0.193  &  1.908  &  ...\\
WISEJ081851.3$+$174326     &  124.71409  &   17.72396 &  11.021  &  0.7834653  & 0.00001    &  105  &  0.282  &  0.276 &           EW &  0.146  &  2.125  &  ...\\
WISEJ073939.5$+$611635     &  114.91497  &   61.27659 &  10.746  &  0.8453609  & 0.00188    &  158  &  0.304  &  0.296 &           EW &  0.115  &  2.095  &  ...\\
WISEJ174404.2$+$383837     &  266.01772  &   38.64386 &  11.511  &  1.0740403  & 0.00208    &  259  &  0.271  &  0.266 &           EW &  0.117  &  2.059  &  ...\\
WISEJ173801.7$+$183017     &  264.50747  &   18.50482 &  11.175  &  0.9878244  & 0.00191    &  178  &  0.213  &  0.223 &           EW &  0.050  &  2.214  &  ...\\
WISEJ174223.8$+$221438     &  265.59954  &   22.24398 &   9.984  &  1.0230755  & 0.00282    &  173  &  0.357  &  0.329 &           EW &  0.200  &  2.039  &  ...\\
WISEJ173400.2$+$161358     &  263.50118  &   16.23280 &   9.794  &  0.9762255  & 0.00188    &  157  &  0.224  &  0.221 &           EW &  0.026  &  1.944  &  ...\\
WISEJ075426.3$+$394659     &  118.60987  &   39.78321 &  11.627  &  1.1033470  & 0.00214    &  110  &  0.405  &  0.386 &     EA/RR/EW &  0.435  &  2.054  &  ...\\
WISEJ181228.7$+$544653     &  273.11978  &   54.78151 &   8.786  &  1.9013039  & 0.00518    &  436  &  0.090  &  0.107 &     EA/RR/EW &  0.330  &  2.204  &  ...\\
WISEJ173313.4$+$065117     &  263.30602  &    6.85493 &  11.627  &  0.9092463  & 0.00248    &  152  &  0.159  &  0.162 &           EW &  0.089  &  2.425  &  ...\\
...                      &  ...        &  ...       &  ...     &  ...        &  ...       &  ...  &  ...    &  ...   &  ...         &  ...    &  ...    &  ...\\
...                      &  ...        &  ...       &  ...     &  ...        &  ...       &  ...  &  ...    &  ...   &  ...         &  ...    &  ...    &  ...\\
...                      &  ...        &  ...       &  ...     &  ...        &  ...       &  ...  &  ...    &  ...   &  ...         &  ...    &  ...    &  ...\\

\hline
\end{tabular}
\tablenotetext{a}{The entire table is available in the online journal;
  50 lines are shown here for guidance regarding its form and
  content.}
\end{center}
\end{table}

\section{The Catalog}

The WISE catalog of variables, containing 50,282 periodic
  variables, is provided in Table \ref{t4}. Fourteen known RRd-type
  variables have been excluded (see Section 5.2 and Table \ref{t3}).
Its column designations represent the following. `ID' refers to the
instrument and position, `SourceID' is the number associated with the
single exposure photometry catalog and `num' is the number of
detections. `$<W1>$, $<W2>$, Period, err$_P/P$, and Amp.$_{\rm{F10}}$'
are the mean magnitudes in two bands, period, period uncertainty, and
amplitude, respectively, determined by means of Fourier fitting. The
other parameters resulting from the Fourier fits, $R_{21}=a_2/a_1$,
$\phi_{21}=\phi_{2}-2\phi_{1}$, $a_4$, and $a_2$ are also
listed. Amp.$_{10}$ is the direct amplitude in the 10--90\% range of
the magnitude distribution (see for details, Section \ref{subamp}) and
`Type' refers to the variable type. The NIR $J, H, \Ks$ magnitudes and
the number of corresponding objects in 2MASS are found under the
column headings `J,' `H,' `$\Ks$,' and `$\rm n_{2MASS}$,'
respectively. The single exposure photometry data and suspected
variables catalog are available online, including their `SourceID,'
`R.A. (J2000),' `Decl. (J2000),' `MJD' (modified Julian date),
`W1mag', `W2mag', `e\_W1mag' and `e\_W2mag'.

\subsection{New variables}

To validate our catalog, a comparison with known variables is
required. The SIMBAD website provides a collection of almost all
variables previously found in the Catalina, OGLE, and ASAS surveys,
among others. We cross matched our 50,296 variables with SIMBAD
positions using a $5''$ angular radius and selected those objects
identified as variables. The majority of the known variables were
found within angular radii of $1-2''$. The variables contained in the
Catalina, OGLE, and ASAS catalogs were supplemented. We found that
15,527 objects had been found previously; the remaining 34,769 are new
variables. Table \ref{t2} lists the statistics of the newly found
variables. EW- and EA-type eclipsing binaries encompass the majority
of the new variables; 1231 new RR Lyrae and 1312 new Cepheids were
also found. Contamination levels were estimated based on comparisons
of our classified types to the known types. For short-period variables
such as EW and RR, the contamination fraction was around 5\%,
increasing to 10\% for long-period variables. In total, 93\% of our
new classifications matched those previously obtained.

\begin{table}[h!]
\vspace{-0.0in}
\begin{center}
\caption{\label{t2}Classification statistics of our 34,769 newly found periodic variables.}
\vspace{0.15in}
\begin{tabular}{lcccc}
\hline
\hline
           Type       &  Count  &  Fraction  &  Purity      &  Contamination                              \\       
\hline                                                                                                         
             EW       &  21427  &   61.63\%  &     95\%     &   RR 2\%, Ro 1\%, EA 1\%                     \\  
             EA       &   5654  &   16.26\%  &     85\%     &   EW 14\%                                  \\    
             RR       &   1231  &    3.54\%  &     93\%     &   EW 5\% EA 1\%                             \\   
       EA/RR/EW       &   1323  &    3.81\%  &   71/16/12\% &                                           \\     
          EW/EA       &   3036  &    8.73\%  &    65/34\%   &   RR 1\%                                   \\    
           Cep-I      &    583  &    1.68\%  &     89\%     &  Cep-II 3\%, EA 3\%, EW 2\%, LPV 2\%, Ro 1\%   \\
     Cep-I/Cep-II     &    492  &    1.42\%  &    57/25\%   &   LPV 8\%, EA 3\%, EW 3\%, Ro 3\%             \\ 
     Cep-II/ACep/Cep-I&    173  &    0.50\%  &   40/29/21\% &   EA 3\%, EW 3\%, Ro 2\%                     \\  
         Cep-I/EA     &     64  &    0.18\%  &   74/22\%    &   Cep-II 4\%                               \\    
           Misc       &    786  &    2.26\%  &              &                                           \\     
          total       &  34769  &    100\%   &              &                                           \\     
\hline
\end{tabular}
\end{center}
\end{table}

\begin{figure}[h!]
\centering
\vspace{-0.0in}
\includegraphics[angle=0,width=180mm]{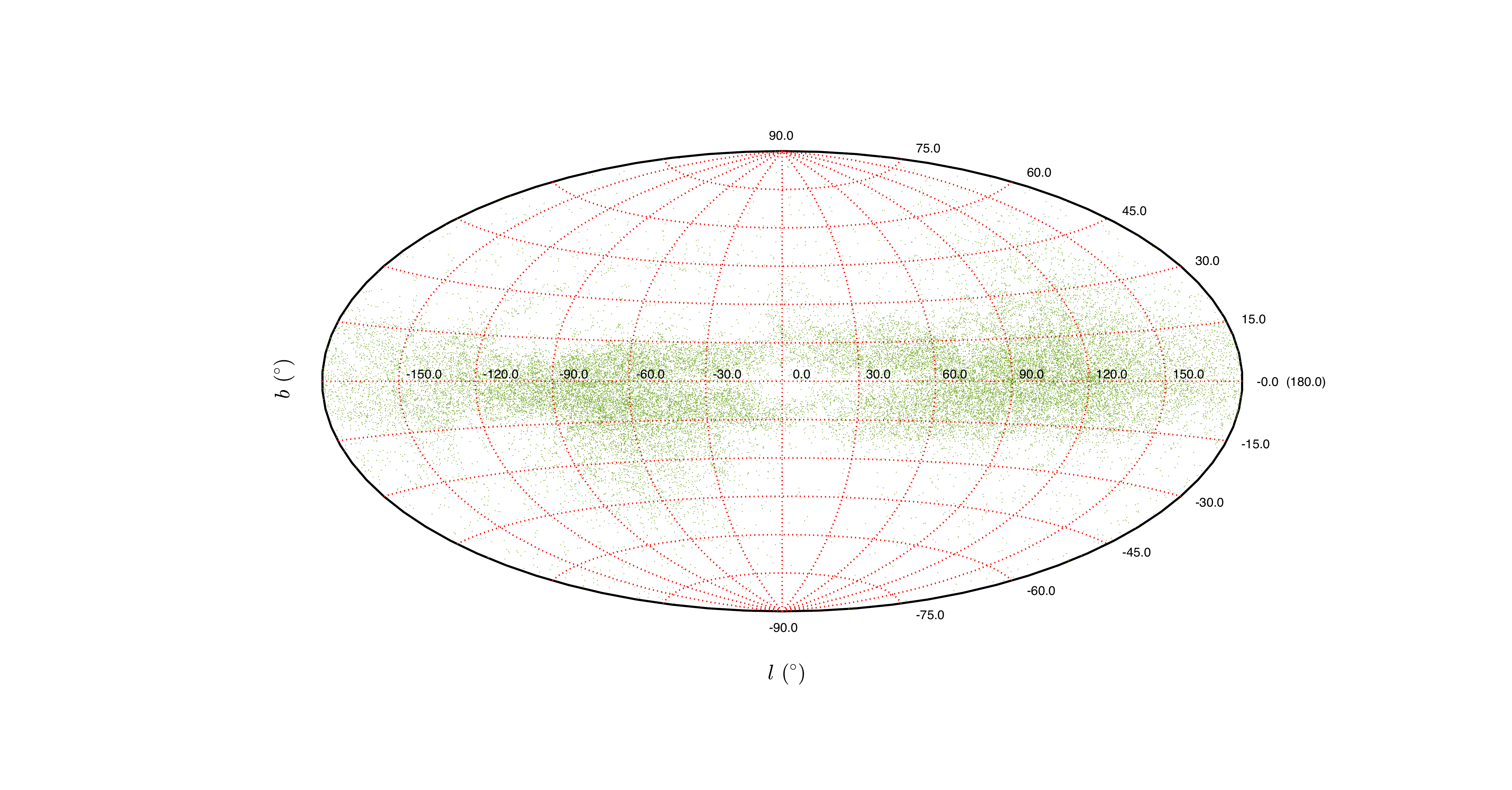}
\vspace{-0.0in}
\caption{\label{map} Distribution of our newly found variables in
  Galactic coordinates. }
\end{figure}

\begin{figure}[h!]
\centering
\vspace{-0.0in}
\includegraphics[angle=0,width=150mm]{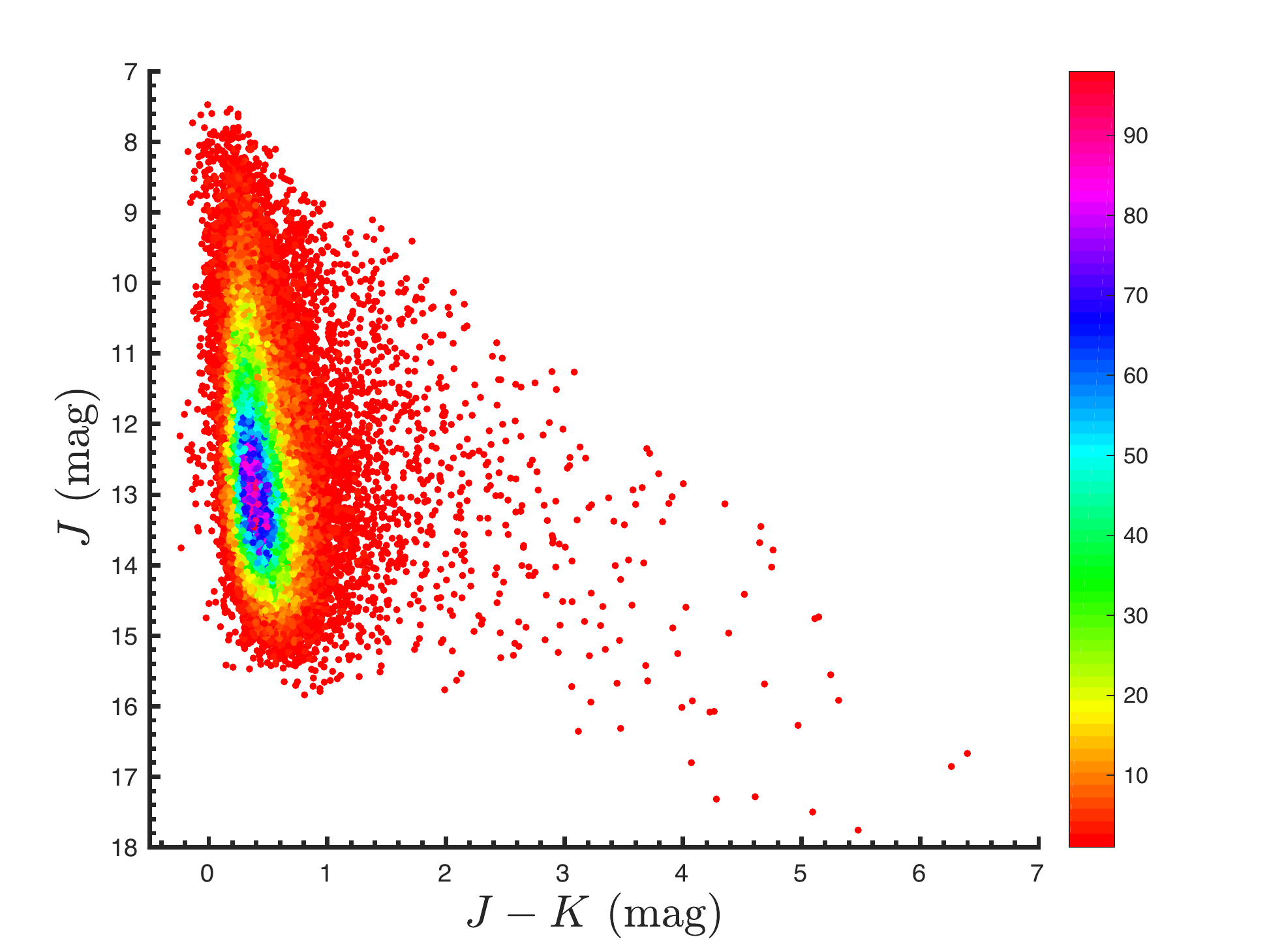}
\vspace{-0.0in}
\caption{\label{cmd} Scatter-density distribution in the NIR
  color--magnitude diagram. The color bar represents the number of
  objects per bin. }
\end{figure}

Figure \ref{map} shows the distribution of the new variables in
Galactic coordinates; 90\% are located within $\pm 20^\circ$ of the
plane, an area obscured in optical surveys. Most of the other new
objects are located close to the equatorial poles, regions that were
poorly sampled by the Catalina survey. A hole is present at the
position of the LMC, where numerous variables were detected by the
OGLE survey. The direction to the Galactic Center is also less
populated, since the WISE detection efficiency decreases in crowded
environments. Figure \ref{cmd} shows the NIR color--magnitude diagram
of these variables; colors denote the number of objects in bins of
$0.035\times0.055$ mag. Approximately 1000 objects have color excesses
$E(J-K)>1.0$ mag, which implies that they are affected by $A_\Ks>0.5$
mag of extinction if we adopt the $A_J/A_{\Ks} =3.13$ NIR extinction
law of \citet{Chen18b}.

\subsection{Comparison with the Catalina catalog}

Compared with SIMBAD, the type classification in the Catalina catalog
is more homogeneous and uniform. Therefore, we also performed
one-to-one matches with the 110,000 variables in the Catalina catalog,
again using a $5''$ angular radius for completeness, of which more
than 10,000 objects matched. Table \ref{t3} reflects the detailed type
comparison. The fraction of misclassifications is similar to that
listed in Table \ref{t2}. For EW-type variables, common contamination
sources are RRc and rotating stars, since both types exhibit symmetric
light curves. Multiple-period RRd-type variables comprise other
  types of contaminants. We have excluded them from our final sample,
  since their main periods were not properly determined in our
  analysis. For EA-type variables, EWs are the main contaminants. The
same conclusion was drawn based on matching the Catalina and LINEAR
catalogs \citep[see][Table 6]{Drake14}. In optical bands, the boundary
between RR Lyrae and eclipsing binaries is obvious and the fraction of
misclassifications is less than 0.1\%. However, in the MIR,
approximately 6\% of RR Lyrae are classified as eclipsing
binaries. Some RR Lyrae show strong phase and amplitude modulations;
they are known as RR Lyrae exhibiting the Blazkho effect. The effect
is also visible but not obvious in the MIR, so we do not identify
these objects separately. `EW/EA' and `EA/RR/EW' types indeed include
high fractions of different types of variables, and no further
classification refinements are possible. Most Cepheids in the Catalina
catalog are Cep-IIs and ACeps, since the survey avoids the Galactic
plane. The rate of contamination is a little higher than in the plane
because of the smaller population of Cepheids here.

\begin{table}[h!]
\small
\begin{center}
\caption{\label{t3}WISE Variables matched with the Catalina catalog}
\begin{tabular}{lccccccccccc}
\hline
\hline
                         
Type           &  EW   & EA  &RRab &RRc &RRd  &Blazkho  &Cep-II&ACep&Rot. & $\delta$ Scuti   &RS CVn \\     
\hline                                                                                                     
EW	           & 7437	 &  28	&   29&	  84&	 12&	    1	&      2	&    0	&  125	&   5	   &    10    \\   
EA	           &  106	 & 640	&    0&	   0&	  0&	    0	&      5	&    1	&  3	  &   0	   &    1     \\   
RR  	         &   57	 &   2	& 1114&	  10&	  0&	    33	&      1	&    5	&  6	  &   0	   &    1     \\ 
EW/EA	         &  611	 & 218	&    0&	   2&	  1&	    0	&      0	&    0	&  1	  &   1	   &    0     \\   
EA/RR/EW	     &   38	 & 117	&   63&	   0&	  0&	    1	&      0	&    0	&  1	  &   0	   &    0     \\ 
Cep            & 3 &	0	&1	&0	&1	&0	&16	&14	&2	&0	&2 \\
\hline
\end{tabular}
\end{center}
\end{table}

\begin{figure}[h!]
\centering
\vspace{-0.0in}
\includegraphics[angle=0,width=150mm]{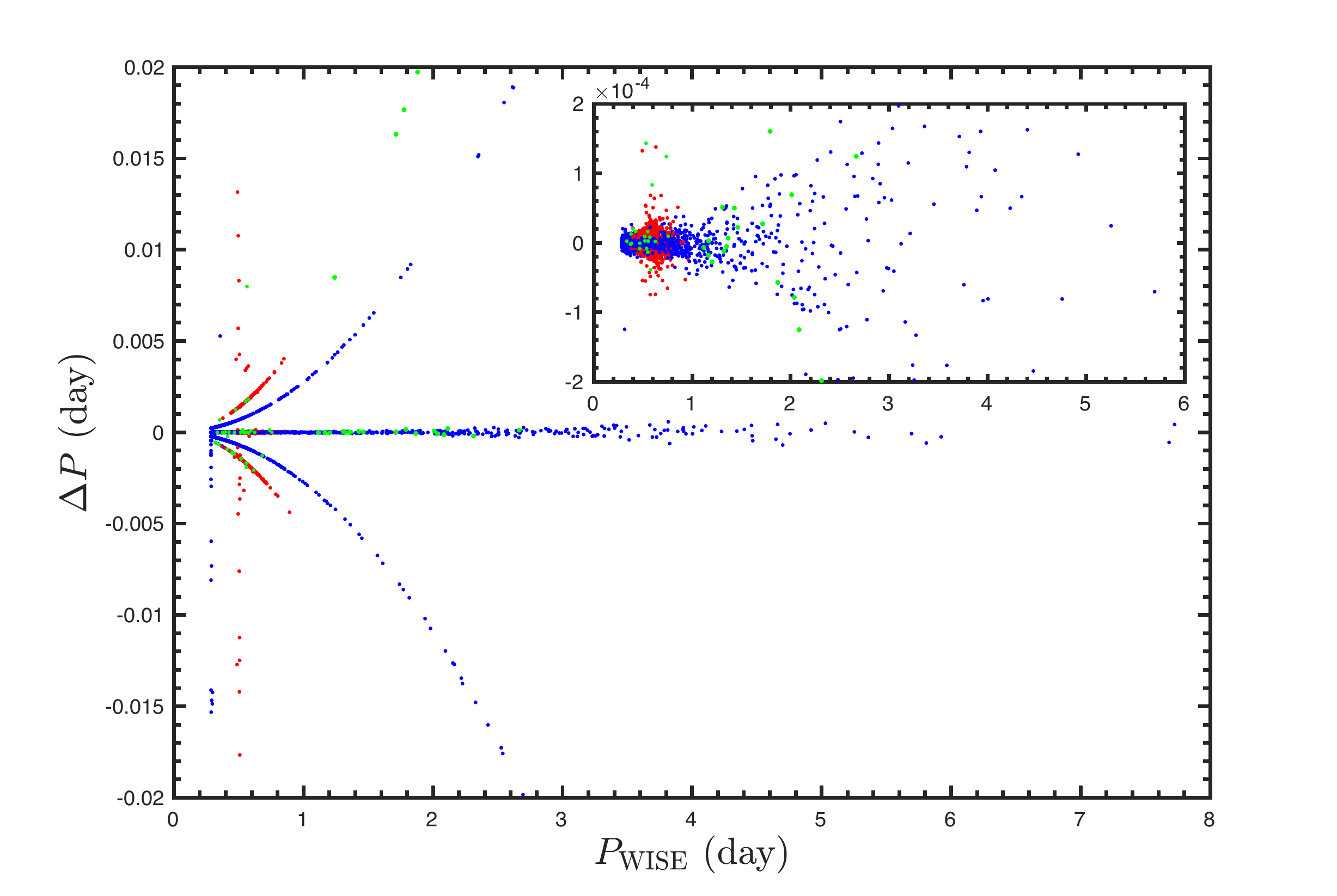}
\vspace{-0.0in}
\caption{\label{per} Period comparison with the Catalina catalog and
  $\Delta P=P_{\rm{Catalina}}-P_{\rm{WISE}}$. The blue, red, and green
  dots are eclipsing binaries, RRab, and other variables,
  respectively. Note that 92\% of the variables are located on the
  central zero line. The top right-hand panel is a zoomed-in view of
  the variables characterized by $-0.0002<\Delta P<0.0002$.}
\end{figure}

We also compared our independently determined periods to those
provided in the Catalina catalog. 98\% of the variables have period
differences $|\Delta P|<0.02$ days: see Figure \ref{per}. The other
2\% of variables showing large differences are mainly eclipsing
binaries. Their period difference is likely driven by the fact that
the harmonic frequency or an adjusted frequency are adopted in the
Catalina catalog. 92\% of the variables have period differences of
less than 0.001 day, and 76\% have period differences of less than
0.0002 day. This implies that our MIR periods are as good as the
optical periods.

\begin{table}[h!]
\begin{center}
\caption{\label{t5}Completeness of WISE variables comparing to Catalina variables}
\begin{tabular}{lcc}
\hline
\hline
Vmag (mag) & Completeness & Completeness (corrected)   \\ 
\hline  
$<11$   &    62.4\%     &    91.4\%     \\   
$<12$   &    57.5\%     &    80.4\%     \\   
$<13$   &    58.4\%     &    82.0\%     \\   
$<14$   &    50.3\%     &    81.5\%     \\   
$<15$   &    38.2\%     &    78.6\%     \\   
$<16$   &    25.9\%     &    74.2\%     \\   
                         
\hline
\end{tabular}
\end{center}
\end{table}

\subsection{Completeness}

 We also compared the sample of known Catalina variables with our
  rediscovered variables sample to evaluate the completeness of our
  catalog. Variables with amplitudes larger than 0.1 mag and periods
  in the range $0.28<P<15$ days are considered, since beyond these
  constraints few variables are detected in our catalog. The
  completeness for each magnitude range is listed in the second column
  of Table \ref{t5}; its level decreases as the variables become
  fainter. In addition, variables with $V>16$ mag are more incomplete
  owing to the detection limit. The $\sim$50\% completeness levels
  mean that our variables are highly affected by our selection
  procedure, which could be improved based on enhanced photometric
  data or in-depth analysis of the poorest-quality light curves. The
  corrected completeness levels are listed in the third column, for
  which we considered suspected variables, non-detections, and
  low-variability-flag objects in the ALLWISE database. An 80\%
  corrected completeness means that the variables' identification
  process is good. The corrected completeness is better than 85\% for
  all variables except for EA-type eclipsing binaries (60\%). This is
  caused by the short time spent in the eclipse phase for EA-type
  eclipsing binaries, which may not be well-detected by WISE.

\subsection{Amplitude comparison}\label{subamp}

\begin{figure}[h!]
\centering
\vspace{-0.0in}
\includegraphics[angle=0,width=150mm]{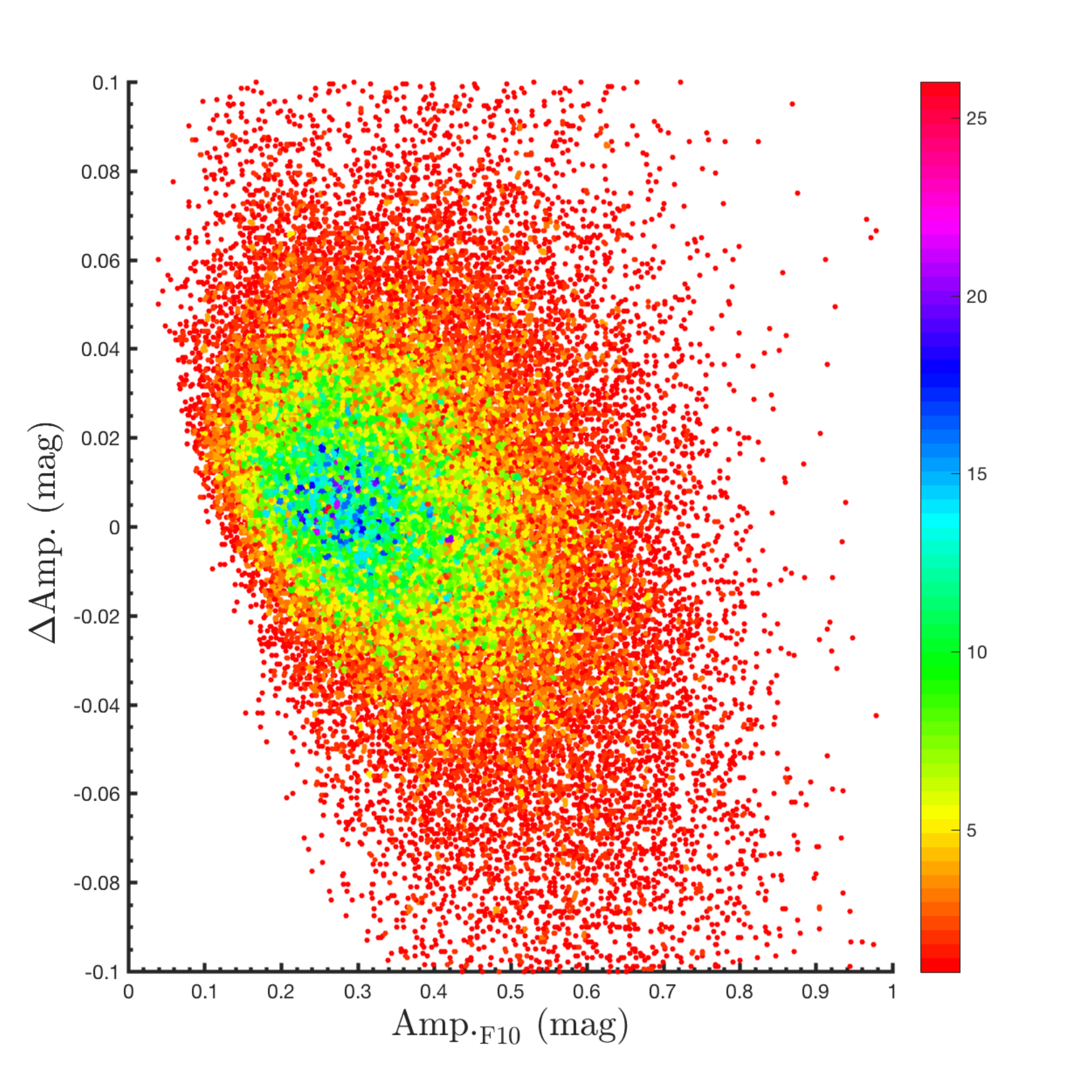}
\vspace{-0.0in}
\caption{\label{amp1} Comparison of the amplitudes determined using a
  10$^{\rm th}$-order Fourier fit, $\rm{Amp.}_{F10}$, and the 10--90\%
  range in magnitude, $\rm{Amp.}_{10}$; $\Delta \rm{Amp.}=
  \rm{Amp.}_{10} - \rm{Amp.}_{F10}$.}
\end{figure}

The amplitudes of the light curves can be estimated based on the best
Fourier fit or directly from the difference between the $n$\% and
$100-n$\% magnitude range. Note that using different methods or
parameters may lead to the determination of different amplitudes. To
determine the most suitable amplitude for our MIR light curves, we
estimated the amplitudes in a number of ways, by varying our input
parameters. The amplitude increases with increasing Fourier-series
order for the first method, while it decreases for increasing $n$ when
applying the second method. We compared the amplitudes resulting from
the two methods and found that the most appropriate amplitudes were
given by the combination of the 10$^{\rm th}$-order Fourier series,
$\rm{Amp.}_{F10}$, and the 10--90\% range in magnitude,
$\rm{Amp.}_{10}$. Figure \ref{amp1} shows the comparison of the two
amplitudes. For variables with small amplitudes ($\rm{Amp.}_{F10}<0.3$
mag) the directly measured amplitude is larger than that from the
Fourier fit. However, for large-amplitude variables
($\rm{Amp.}_{F10}>0.3$ mag), the best-fitting Fourier amplitude is the
larger one. We have included both amplitudes in the final catalog.

\begin{figure}[h!]
\centering
\vspace{-0.0in}
\includegraphics[angle=0,width=150mm]{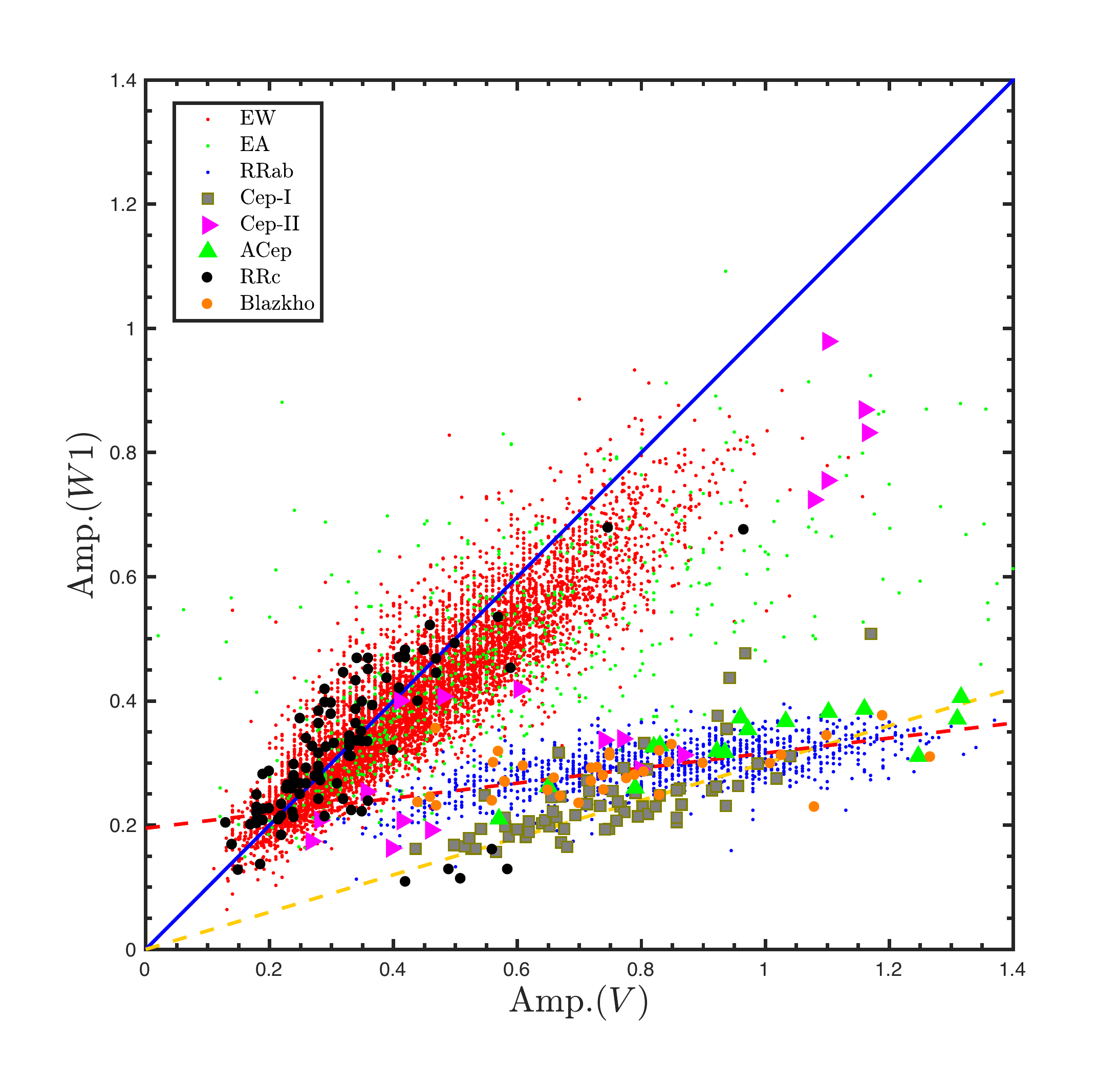}
\vspace{-0.0in}
\caption{\label{amp2} Comparison of the $W1$ and $V$-band amplitudes
  for different variables. Red dots: EW; green dots: EA; blue dots:
  RRab; grey squares: Cep-I; magenta right-pointing triangles: Cep-II;
  green upward triangles: ACep; black solid circles: RRc; orange solid
  circles: Blazkho RR Lyrae. The blue line is the one-to-one line,
  while the red and yellow dashed lines are the best-fitting lines for
  RRab and Cep-I, following $A_{W1}=(0.121\pm0.004) A_{V} +
  (0.195\pm0.003)$ mag and $A_{W1}=(0.30\pm0.01)A_{V}$ mag,
  respectively.}
\end{figure}

Figure \ref{amp2} shows a comparison of $W1$ and $V$-band amplitudes;
best-fitting Fourier amplitudes ($\rm{Amp.}_{F10}$) have been
adopted. The EW, EA, RRab, and Cep-II variables have the same
classifications in both our catalog and the Catalina catalog. RRc and
Blazkho RR Lyrae have not been classified separately in our catalog,
so we adopted the Catalina types here. Cep-Is were compared with the
446 classical Cepheids studied by \citet{Berdnikov08}. Our comparison
shows that amplitude changes caused by geometry, such as eclipse and
radius variations, follow the one-to-one line, while amplitude
modulation caused by effective temperature variations follow a much
shallower trend (see the dashed lines). The two components of EWs have
similar temperatures, and the amplitude modulation is mainly caused by
eclipses, which result in similar amplitude from the optical to the
infrared regime \citep[see also][]{Chen16}. For EA-type variables,
both components have different effective temperatures, and they
therefore deviate from the one-to-one line.

Pulsating stars (in their fundamental mode) exhibit obvious effective
temperature variations, and they follow the dashed lines. In detail,
RRab and Cep-Is follow different relations, $A_{W1}=(0.121\pm0.004)
A_{V} + (0.195\pm0.003)$ mag and $A_{W1}=(0.30\pm0.01)A_{V}$ mag,
respectively. These are comparable to previously published NIR
template light curves: RRab, $A_{K}=(0.108\pm0.018) A_{B} +
(0.168\pm0.024)$ mag \citep{Jones96}; Cep-Is, $A_{\Ks}=(0.32\pm0.01)
A_{V}$ mag \citep{Inno15}. Orange solid circles are RR Lyrae
exhibiting the Blazkho effect. They follow the same relation as the
RRab. Unlike fundamental-mode pulsating stars, overtone pulsators have
similar amplitudes in optical and infrared bands. Black solid circles
represent RRc; they are close to the one-to-one line. Eight LMC
first-overtone Cepheids (not shown) also have similar $I$- and
$W1$-band amplitudes. Cep-IIs are more complicated; long-period
Cep-IIs ($P>10$ days) are located close to the one-to-one line, while
short-period Cep-IIs are located close to pulsating line, with some
exceptions. These features are also visible in the NIR bands
\citep[Figure 1]{Bhardwaj17}. This diagram provides another means to
classify variables when the number of detections is insufficient for
light curve analyses.

The good agreement in terms of classification, period, and amplitude
with respect to the Catalina catalog validates that our MIR variable
characterization is indeed highly reliable.

\section{Light curves and applications}

In this section, some light curves of the new variables are shown and
a brief discussion of their possible future application is presented.

\subsection{Cepheid}
\begin{figure}[h!]
\centering
\hspace{0.0in}
\includegraphics[angle=0,width=180mm]{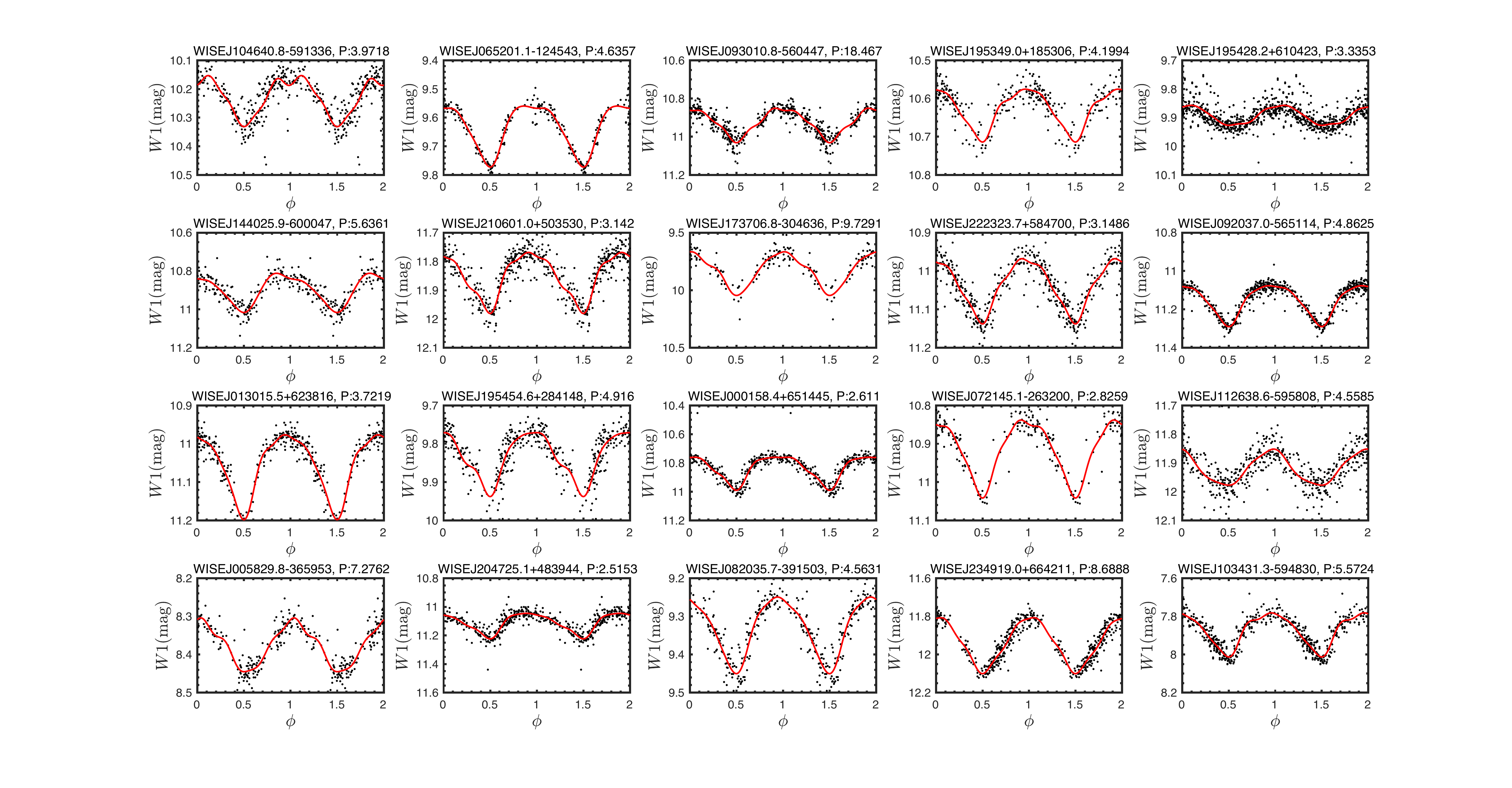}
\vspace{-0.0in}
\caption{\label{cep} Light curve examples for 20 of the new
  Cepheids. The black dots and red line are the observed and
  best-fitting light curves, respectively. The full set of light
  curves is available online. }
\end{figure}

Classical Cepheids are the most important and among the most accurate
distance indicators used for establishing the astronomical distance
scale. Compared with extragalactic Cepheids, Cepheids in our Galaxy
are poorly detected. Some 500 of Cepheids with good photometry were
listed by \citet{Berdnikov08}. Around 800 Cep-Is are included in the
ASAS catalog \citep{Pojmanski05}. However, contamination of the sample
is a problem. To date, only 450 Cepheids are usually used to study
Galactic structure \citep{Genovali14}. Note that these Cepheids are
predominantly located in the solar neighborhood. Because of the heavy
reddening and crowded environment in the inner plane, only a few to a
few dozen Cepheids have been found there \citep[e.g.][]{Matsunaga13,
  Matsunaga15, Dekany15, Tanioka17, Inno18}. Here, we report 1312
newly discovered Cepheids; the majority are classical Cepheids in the
inner plane. This should promote a significantly better understanding
of the inner plane's structure. The huge increase of Galactic Cepheids
also offers the possibility to better constrain the scatter and
zeropoint of the Cepheid period--luminosity relation (PLR). In
addition, Cepheids are usually associated with young open clusters or
OB associations. Finding new open cluster Cepheids is valuable
\citep{Chen15, Chen17, Lohr18}, since it may help us understand the
evolution of intermediate-mass stars \citep{Smiljanic18}. Figure
\ref{cep} shows 20 randomly selected Cepheid light curves. Our Cepheid
sample is contaminated by roughly 10\% of objects that are actually
eclipsing binaries, rotating stars, and long-period quasi-periodic
variables; their nature should be double checked based on their color,
extinction, and distance properties.

\subsection{RR Lyrae}

\begin{figure}[h!]
\centering
\hspace{0.0in}
\includegraphics[angle=0,width=180mm]{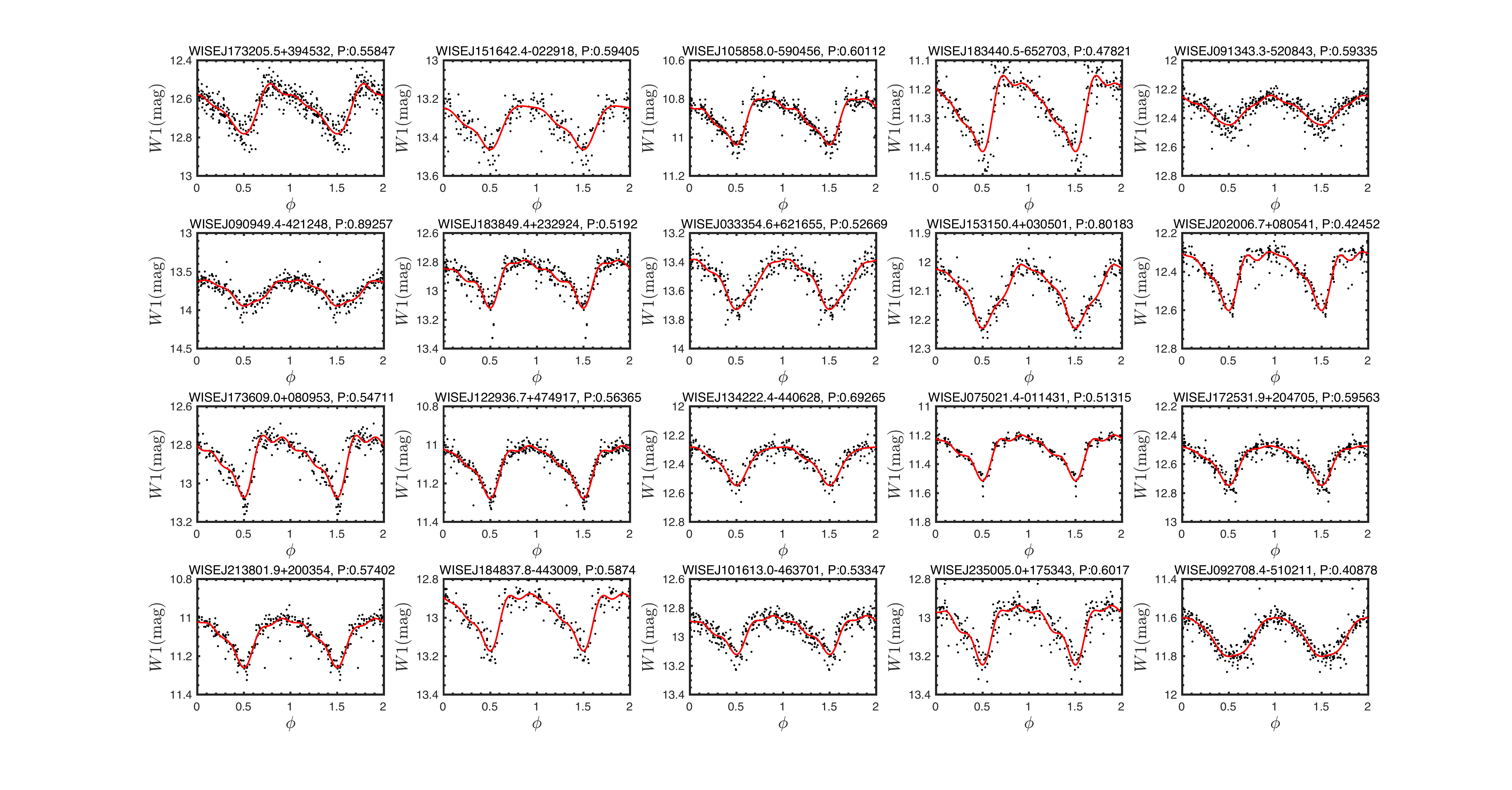}
\vspace{-0.0in}
\caption{\label{RR} Example light curves of 20 new RR Lyreas. The
  black dots and red line are the observed and best-fitting light
  curves, respectively. The full set of light curves is available
  online.}
\end{figure}

RR Lyrae comprise another useful distance indicator that could trace
old environments with 4--5\% distance accuracy \citep[and references
  within]{Catelan09}. They are advantageous to study the Galactic halo
\citep{Vivas06, Sesar10, Drake13}, bulge \citep{Gran16}, as well as
the solar neighborhood \citep{Layden94, Layden98}. The 1231 newly
discovered RR Lyrae in our catalog expand the number of known RR Lyrae
in the solar neighborhood, especially in reddened regions. Figure
\ref{RR} shows example light curves of new RR Lyrae. In this period
range, 6\% of misidentified eclipsing binaries are hard to exclude
only based on their light curves. Colors, independent distances, or
amplitudes in optical bands would be helpful to improve the sample's
purity.

\subsection{Eclipsing binaries}

\begin{figure}[h!]
\centering
\hspace{0.0in}
\includegraphics[angle=0,width=180mm]{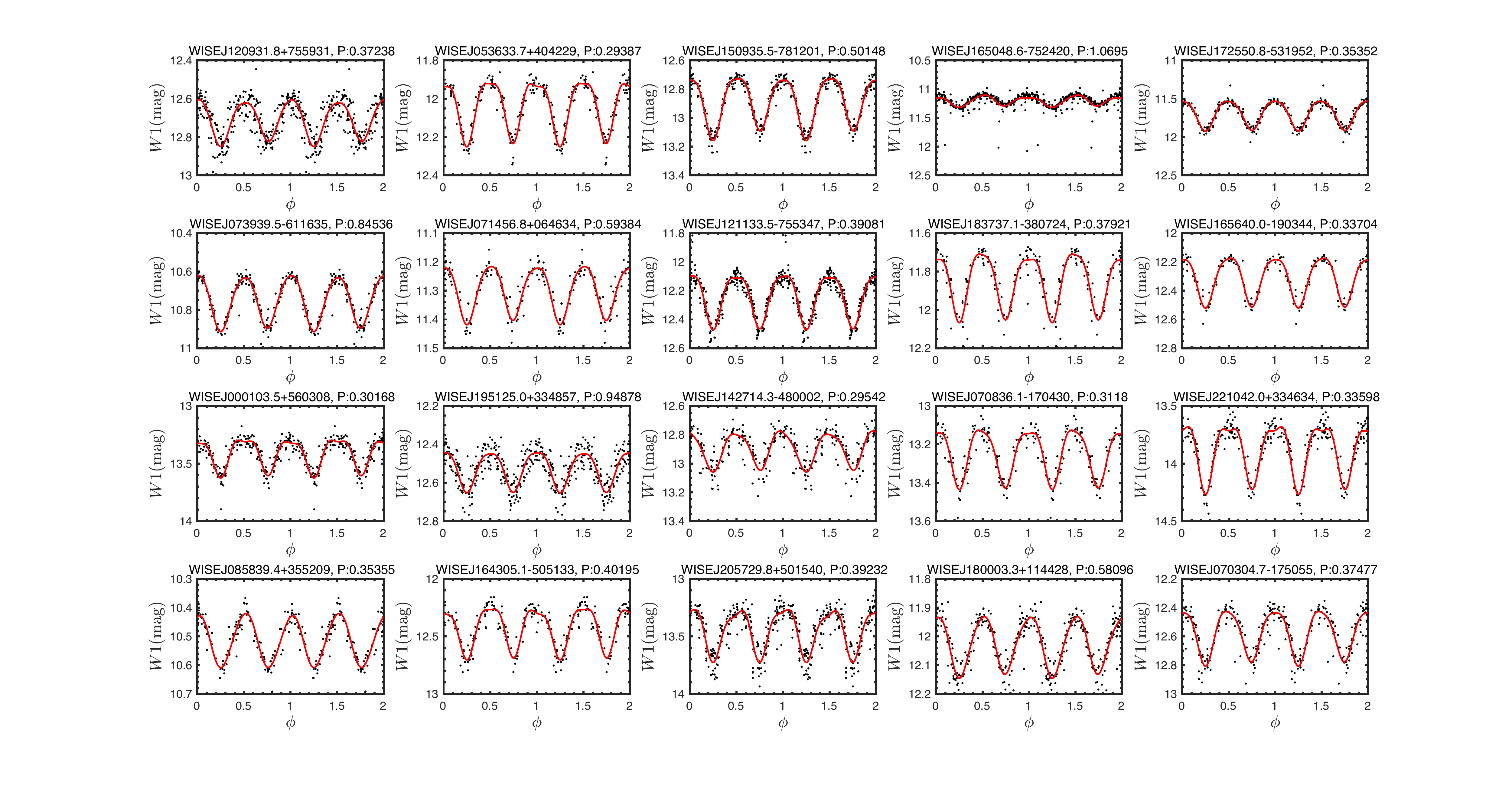}
\vspace{-0.0in}
\caption{\label{EW} Example light curves of 20 new EW. The black dots
  and red line are the observed and best-fitting light curves,
  respectively. The full set of light curves is available online.}
\end{figure}

Eclipsing binaries are among the most plentiful variables in the
periodic variable catalog. Based on their light curves, they are
divided into three subtypes: EA, EB, and EW type refer to detached,
semi-detached, and contact binaries. The majority of EW types are
over-contact, contact, or near-contact eclipsing binaries, which
follow the relevant PLR in some period range. In particular, W
UMa-type contact binaries in the period range 0.25--0.56 days are
important distance indicators. \citet{Chen18a} established their PLRs
based on {\sl Gaia} DR1 parallaxes \citep{Gaia16}. They could be used
to trace distances to an accuracy of 7--8\% in $W1$ band. With the
updated {\sl Gaia} DR2 parallaxes, the accuracy could be improved to
6\% (0.13 mag) in the $W1$ band. \citet{Chen18a} also studied the
structure of the solar neighborhood based on more than 20,000 W
UMa-type contact binaries from the Catalina and ASAS
catalogs. However, the Galactic plane was not covered. The 20,000
newly discovered EW-type eclipsing binaries fill in the Galactic plane
and allow us to construct a complete sample covering distances out to
2--3 kpc from the Sun. Figure \ref{EW} shows example light curves of
EW-type eclipsing binaries.

\begin{figure}[h!]
\centering
\hspace{0.0in}
\includegraphics[angle=0,width=180mm]{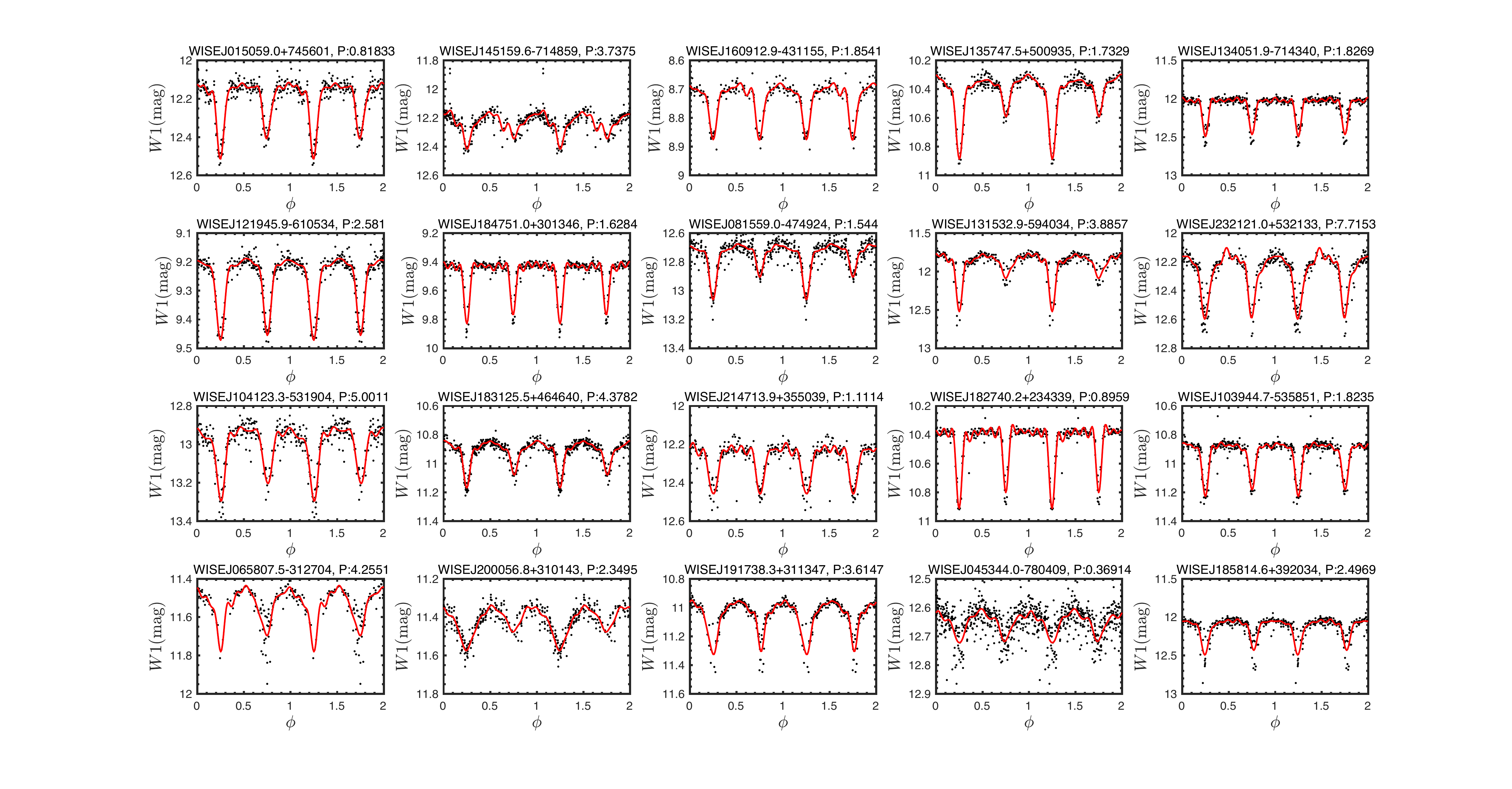}
\vspace{-0.0in}
\caption{\label{EA} Example light curves of 20 new EA-type
  binaries. The black dots and red line are the observed and
  best-fitting light curves. The full set of light curves is available
  online. }
\end{figure}

Compared to EW, it is easy to identify the eclipse onset in EA light
curves (see Figure \ref{EA}). The majority of EAs are detached
eclipsing binaries. Combining the light and radial velocity curves,
detached eclipsing binaries are the best objects one can use to
determine stellar parameters and accurate distances \citep{Grundahl08,
  Pietrzynski13}.

\section{Conclusions}
In this paper, we have collected five-year ALLWISE and NEOWISE-R data
to detect periodic variables. Based on 2.7 million high-probability
candidates in ALLWISE, we found 50,282 periodic variables and an
additional 17,000 suspected variables. With more than 100 detections
each, these variables represent a well-observed catalog. A global
variable classification scheme based on $W1$-band light curves was
established and the first all-sky infrared variable star catalog was
constructed. Among the catalog's 50,282 periodic variables, 34,769 are
newly discovered, including 21,427 EW-type eclipsing binaries, 5654
EA-type eclipsing binaries, 1312 Cepheids, and 1231 RR Lyrae. The
newly found variables are located in the Galactic plane and the
equatorial poles, which where were not well covered in earlier
studies. Since the $W1$-band extinction is one-twentieth of that in
the $V$ band, these variables can be used to pierce through the heavy
dust in the Galactic plane. A careful parameter double check with the
literature for the reconfirmed variables was done, resulting in
misclassification rates of 5\% and 10\%, respectively, for short- and
long-period variables. Type, period, and amplitude comparisons with
the Catalina catalog not only validated our sample of variables, but
also implied different optical--infrared properties for different
types of variables.

The newly found variables can be used to trace new structures or
better study the Galactic plane, the spiral arms, and the solar
neighborhood. Especially for Cepheids, the number of known objects has
increased by an order of magnitude in the inner plane. The scatter,
zero point, and systematics of Cepheid, RR Lyrae, and W UMa-type
contact-binary PLRs will also benefit from this expanded sample. These
objects are helpful to study stellar evolution based on a more
complete sample. In addition, based on some variables located in
heavily reddened regions, the infrared extinction law may be better
characterized. To avoid the detection of false positives, these
  50,282 variables are highly selected variables with completeness
  levels in excess of 50\%. WISE operations are continuing, and an
  additional 30,000 variables are expected to be found based on the
  enhanced data. These MIR variables are also interesting candidates
for future targeted campaigns with the {\sl James Webb Space
  Telescope}.

\acknowledgments{We thank the anonymous referee for suggestions to
  help us improve the paper. This publication makes use of data
  products from ALLWISE and NEOWISE, which are projects of the Jet
  Propulsion Laboratory/California Institute of Technology. ALLWISE
  and NEOWISE are funded by the National Aeronautics and Space
  Administration. We are grateful for research support from the
  National Natural Science Foundation of China through grants
  U1631102, 11373010, and 11633005, from the Initiative Postdocs
  Support Program (No. BX201600002), and from the China Postdoctoral
  Science Foundation (grant 2017M610998). This work was also supported
  by the National Key Research and Development Program of China
  through grant 2017YFA0402702.}

\end{document}